\PassOptionsToClass{journal,10pt,twocolumn,letterpaper}{IEEEtran}

\documentclass[journal]{IEEEtran}
\usepackage[T1]{fontenc}
\usepackage{amsmath}
\usepackage{amssymb}
\usepackage{cite}
\usepackage{graphicx}
\usepackage{xcolor}
\usepackage{dblfloatfix}
\usepackage{float}
\usepackage{booktabs}
\usepackage{tabularx}
\usepackage{listings}
\usepackage{etoolbox}
\usepackage[hyphens]{xurl}
\usepackage[multiple]{footmisc}

\newtoggle{debugmode}

\togglefalse{debugmode}

\iftoggle{debugmode}{%
  \usepackage{hyperref}
  \usepackage[expansion=false]{microtype}
  \setlength{\overfullrule}{5pt}
  \usepackage{layout}
  \usepackage{showframe}
}{%
  \usepackage[hidelinks]{hyperref}
}

\hypersetup{unicode=true}
\graphicspath{{content/article/figures/}}
\newcommand{\software}{gdelt-regkg}

\newcommand{\github}{\url{https://github.com/octo-gone/gdelt-regkg/tree/v2.2.0}}
\newcommand{\pypi}{\url{https://pypi.org/project/gdelt-regkg/2.2.0}}

\let\oldFootnote\footnote
\newcommand\nextToken\relax

\renewcommand\footnote[1]{%
    \oldFootnote{#1}\futurelet\nextToken\isFootnote}

\newcommand\isFootnote{%
    \ifx\footnote\nextToken\textsuperscript{,}\fi}
\newcommand{\ArticleNote}[1]{%
  \par\smallskip\begingroup
  \footnotesize\raggedright
  \noindent#1\par
  \endgroup
}

\newcommand{\ArticleReferences}{%
  \bibliographystyle{styles/IEEEtran-extended}%
  \bibliography{content/references/references}%
}

\title{ReGKG-2025: A GKG-Compatible Reconstruction of the GDELT Global
Knowledge Graph for the Missing Period of 14 June to 2 July 2025}
\author{Nikita~A.~Klimin%
\thanks{N.~A.~Klimin is with HSE University, Moscow, Russia (e-mail: naklimin@edu.hse.ru).}%
}

\newcommand{\ArticleIndexTerms}{Dataset reconstruction, GDELT,
Global Knowledge Graph, news archives, source coverage, temporal validation.}

\begin{document}
\maketitle

\begin{abstract}
    The June 2025 outage in the Global Database of Events, Language, and
Tone (GDELT) collection pipeline left a gap in the Global Knowledge
Graph (GKG), interrupting article-level news metadata used in downstream
research. ReGKG-2025 is an archive-derived sample of 281,620
reconstructed article records covering all 1,665 missing 15-minute
batches. Articles were recovered from Common Crawl and the Wayback
Machine using source quotas estimated from surrounding native GKG data.
Text was screened and normalized, non-English bodies were translated,
and flagged records were manually reviewed. Eligible articles were
converted to the 27-column GKG 2.1 layout. The versioned package
includes record-level provenance, archive replay and WARC locators,
source summaries, source-by-day calibration weights, and the complete
manual decision journal. The sample reaches about 14\% of modeled native
volume and represents sources accounting for 84\% of selected reference
record mass. In an intact-week test, 41\% of recovered articles matched
native GKG records; field agreement on matched pairs is close to the
processor baseline (tone MAE 1.69). The dataset supports retrospective
content analysis and source-level aggregates; its tone and theme
distributions differ from native GKG, and it does not support exhaustive
counts. Data: \url{https://doi.org/10.5281/zenodo.23964265}.

\end{abstract}

\section{Introduction}\label{sec:introduction}
\IEEEPARstart{D}{uring} the second half of June 2025, the Global
Database of Events, Language, and Tone (GDELT) experienced a
data-collection outage. The interruption presumably followed the global
Google Cloud Platform Service Control failure on 12 June, which
disrupted API access across Google Cloud
services~\cite{google2025servicecontrol}. Downstream users reported
interrupted news monitoring, for example, the University of Maryland's
Cyber Events Database states that the outage reduced its identification
of events reported during the second half of June~\cite{umdcyberevents}.
The official interval-file inventory for Global Knowledge Graph (GKG),
one of GDELT's main databases, identifies 1,665 consecutive missing
15-minute buckets from 14 June 2025, 18:00, through 2 July 2025, 02:00
UTC, inclusive~\cite{gdelt2026masterfilelist}.

The GKG itself provides processed news articles as extracted themes,
entities, locations, tone, and other content
measures~\cite{leetaru2015gkgcodebook}. Because GKG is widely used,
temporal gaps in these records can disrupt downstream data collection,
processing, and analysis.

One possible response is to reconstruct missing GKG records from
archived news. To address this, we developed Python scripts and later
released them as the gdelt-regkg package~\cite{klimin2026gdeltregkg}.
Using this package we created the ReGKG-2025
dataset~\cite{klimin2026regkgdataset}. It contains an archive-derived
sample for the missing interval: 281,620 GKG-compatible article records
from 4,301 sources covering all 1,665 buckets. Recovery reaches
approximately 14\% of the original volume (estimated from surrounding
native weeks). Represented sources contain 77.52\% of selected source
identities and account for 83.94\% of selected reference record mass;
these are coverage measures with different denominators
(Table~\ref{tab:dataset-glance}).

\begin{table}[!th]
  \caption{Dataset at a Glance}
  \label{tab:dataset-glance}
  \centering
  \footnotesize
  \setlength{\tabcolsep}{3pt}
  \begin{tabular*}{\columnwidth}{@{\extracolsep{\fill}}lr@{}}
    \toprule
    Property & Value \\
    \midrule
    Version & 1.0.0 \\
    Article records & 281,620 \\
    Represented sources & 4,301 \\
    Occupied / target batches & 1,665 / 1,665 \\
    Selected-source identities & 77.52\% \\
    Selected reference mass covered & 83.94\% \\
    Attained modeled volume & 13.63--14.28\% \\
    License & CC BY 4.0 \\
    \bottomrule
  \end{tabular*}
  \ArticleNote{The volume range uses the full and selected source frames. Source presence, reference mass, and attained article volume are different quantities.}
\end{table}

This work provides:
\begin{itemize}
  \item a GKG-compatible reconstruction sample covering every missing
  batch in the outage interval;
  \item record and archive provenance, review decisions, and source/day
  weights for selecting and adjusting analysis subsets;
  \item validation of timing, source and content distributions, and
  paired field agreement through an intact-week reconstruction.
\end{itemize}

The GKG layout and the source/day weights allow existing pipelines to
ingest the recovered records alongside native data, subject to the
limitations described in Section \ref{sec:limitations}.

\section{Background}\label{sec:related-work}
Article retrieval is typically the most resource-intensive step in
building a news dataset. Possible sources include pre-existing datasets,
news services, archived web pages, and publisher APIs, but their
suitability depends on whether they provide both full article text and
publication metadata for the sources and date range.

\subsection{Access to News Articles}
\label{sec:related-access}
A prepared dataset avoids repeating part of the collection and
extraction work, but in most cases its selection was designed for a
different research question. NELA-GT-2018, for example, supports
misinformation research through a collection of selected outlets that
can bias results toward those outlets~\cite{norregaard2019nelagt}. A
closer variant is Fine-News~\cite{solovev2026finenews}, which derives
news from Common Crawl through
Infini-News~\cite{lazzaroni2026infininewsefficientlyqueryableaccess},
with language checks, quality filtering, and duplicate removal. However,
reusing such a corpus requires checking whether its source and date
coverage, available metadata, and filtering decisions are compatible
with the reconstruction of GKG records.

When a prepared collection does not cover the target, news services
allow more specific requests, but the cost of historical access can be a
substantial barrier to large-scale reconstruction, and free plans do not
remove this barrier. Event Registry and NewsData.io meter historical
retrieval through paid usage
allowances~\cite{eventregistryplans,newsdataarchive}. As an
illustration, NewsData.io's listed monthly prices, consulted in October
2026, were US\$349.99 for a plan with two years of historical access and
US\$1,299.99 for ten years, each with a finite credit allowance; archive
requests consumed five credits per call~\cite{newsdatapricing}. Event
Registry also charges more tokens for historical searches, with each
page of results charged separately~\cite{eventregistryplans}. Both
providers advertise academic discounts; NewsData.io lists reductions of
up to 90\%, subject to contacting the
provider~\cite{eventregistryplans,newsdataacademic}. Some datasets
follow the same model: the NOW corpus provides continuing access to
English-language news and distributes bulk text, but requires payment,
for example, at least US\$395 for one corpus and more for additional
content~\cite{nowcorpusdata}. List prices therefore illustrate potential
costs without establishing the price available to a particular research
project. Besides, any acquisition budget can limit the volume and
historical depth of a reconstruction even when the requested articles
are available.

Paying for access also does not guarantee the required coverage or text.
Factiva and LexisNexis provide licensed archives whose usable coverage
depends on the publications and content
included~\cite{proquestfactivacontent,lexisnexisnewsdata}, and access to
articles does not always include the text needed for reprocessing. Media
Cloud supports source discovery and multi-source
analysis~\cite{bermejo2026mediacloud}, but does not release article
bodies through its public tools~\cite{mediacloudfaq}. Publisher APIs
similarly differ in the content they expose. The Guardian Open Platform
offers article text for its own publication~\cite{guardianaccess},
whereas The New York Times Archive API supplies only article
metadata~\cite{nytimesarchiveapi}. Combining such services still
requires checking text availability and reconciling their coverage and
metadata.

Web archives provide a route to historical text outside prepared
collections and service subscriptions. Common Crawl News publishes
webpage captures as WARC files and provides tools to index and request
content from them~\cite{nagel2016ccnews}. Similarly, the Wayback Machine
offers historical page versions~\cite{internetarchive2001wayback} with a
programmatic capture index~\cite{internetarchivecdxapi}. These inputs
leave the work that a prepared corpus has already performed: identifying
relevant captures, retrieving pages, extracting article bodies, and
recovering publication metadata. CC-News-En documents this process for
an English-language corpus derived from Common Crawl
News~\cite{mackenzie2020ccnews}. Archive-based collection can therefore
support reconstruction, with coverage limited to the pages captured and
recoverable from each archive.

This limit is not neutral, because what an archive holds is the result
of selection. Estimated archival coverage changes substantially with the
set of URLs used to represent the
web~\cite{ainsworth2011archived,hale2017livearchive}, and Common Crawl's
crawl priorities and language imbalances shape what it
contains~\cite{baack2024commoncrawl}. Neither study measures recovery
for the interval considered here, but together they show that an
archive's size is not evidence of balanced coverage. A reconstruction
must therefore measure which sources and languages survive its own
discovery and retrieval stages.

The final option, collecting directly from publishers, allows
source-specific retrieval without relying on an earlier archive capture,
but it can only recover what is still online, and in the form it has
now. Previously accessible pages disappear, including from otherwise
functioning websites~\cite{chapekis2024disappears}, and a reachable URL
may serve different content than at publication. For instance,
discrepancies between browsing captures and later scrapes of the same
URLs vary across news categories~\cite{ulloa2025timedelays}. Archived
copies are also affected, since replayed pages can mix archived HTML
with JSON from another date~\cite{weigle2023replay}.

\subsection{From Recovered Pages to Article Records}
\label{sec:related-preparation}
Once pages have been recovered, extraction quality determines which text
is available for analysis, and it is not uniform. Main-content
extractors perform differently depending on page genre and
complexity~\cite{bevendorff2023extraction}, and degrade as website
structures change over time~\cite{weninger2015extraction}. A single
extractor therefore needs checks for boilerplate and missing passages
across publishers and historical layouts.

\subsection{Representativeness and Integration with Native Data}
\label{sec:related-representativeness}
The articles that survive retrieval and preparation determine what the
sample can represent. For archive-based collection this means that
source quotas describe an intended allocation, while archive
availability and retrieval success decide which articles can actually
fill it. The reference distribution is itself limited, because GKG
covers only a subset of news with strong regional
bias~\cite{kwak2014newsgeography}. Matching source shares from GKG
therefore preserves a distribution of observed reporting; it cannot
establish balanced coverage of all news or of all real-world events.

\subsection{Validation of Reconstructed Records}
\label{sec:related-validation}
Validation concerns the articles recovered, the fields generated from
them, and their usefulness in downstream tasks. Sen et al.'s total-error
framework separates representation and measurement errors in digital
traces~\cite{sen2021totalerror}. For article recovery, this means
checking text validity and coverage across sources, languages, and
dates. Braun and Oswald report temporal correspondence with external
document counts alongside undercoverage in an archive-derived
travel-advisory corpus~\cite{braun2026validating}; matching trends alone
therefore does not establish complete recovery.

For generated fields, agreement with native GKG assesses consistency
with an existing coding system, as in the gdelt-regkg
study~\cite{klimin2026agreement}. Assessing whether a field is supported
by its article also requires independent inspection, consistent with
Grimmer and Stewart's emphasis on application-specific validation and
close reading~\cite{grimmer2013text}. Matched article versions allow
paired checks; when the articles differ, source, language, and time
composition can also explain differences between field distributions.

Downstream tests could assess whether the resulting records support a
specific task, such as forecasting. They could provide evidence of
practical utility alongside retrieval and field checks; predictive
performance alone does not establish article coverage or field
correctness. This paper includes descriptive tone comparisons, but does
not conduct dedicated evaluations of downstream tasks.

\section{Methods}\label{sec:methods}
\subsection{Reference Data and Source Allocation}\label{sec:reference-data-and-source-allocation}
We estimated the source distribution from official GDELT Global
Knowledge Graph (GKG) records preceding and following the missing
interval, from 14 June 2025, 18:00, through 2 July 2025, 02:00 UTC,
inclusive. The reference data comprised 21 daily Parquet files
containing GKG records for 1--14 June and 2--8 July 2025. These files
were produced by collecting the original GKG interval files and grouping
their records by calendar day. Malformed fragments were removed before
analysis. The reference contained 2,490,627 records, obtained from
12,459 sources and corresponding to 1,974 out of 2,016 15-minute
intervals (boundary dates are partial).

We counted reference rows for each source using the
\texttt{V2SOURCECOMMONNAME} field. Counts were pooled across both
reference windows without giving each day equal weight. We retained
sources with at least 50 reference rows and at least one dated record on
each side of the gap. The resulting source set, $S$, contained 5,548
sources and 2,377,392 reference rows, accounting for 95.45\% of all
reference rows. The target source weights were
\begin{equation}
  p_s = \frac{n_s}{\sum_{j \in S} n_j},
  \label{eq:source-weights}
\end{equation}
where $p_s$ is the target share of source $s$ and $n_s$ is its reference
GKG row count. We also recorded each source's share of all reference
rows, including sources outside $S$. These weights describe source
frequencies in the reference GKG records.

Sources were identified by GDELT's source names. For example, station
hosts attributed to \texttt{iheart.com} were treated as one source for
allocation, while their individual URLs were used for retrieval. Sources
classified as listings or aggregators were included if they met the same
selection criteria. The reference distribution therefore inherited these
source categories and GDELT's relative source frequencies.

\begin{figure*}[!b]
\centering
\includegraphics[width=\textwidth]{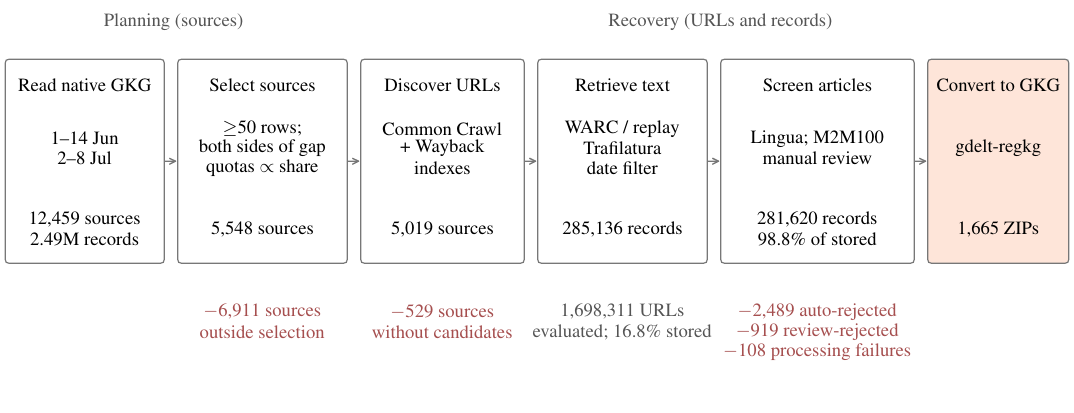}
\caption{Collection pipeline with source and record counts.}
\label{fig:pipeline-funnel}
\end{figure*}

\subsection{Reference Model and Volume Estimates}\label{sec:reference-model}
We modeled native GKG record rates from the surrounding reference weeks
to estimate the volume expected during the missing interval. These
estimates provide common denominators for source allocation,
attained-volume ratios, and temporal comparisons. Because the reference
and reconstruction cover different calendar periods, the raw totals from
the 21 reference dates cannot be used directly as the expected count for
the reconstruction's 1,665 target intervals.

We first grouped the observed native intervals by weekday and 15-minute
UTC clock position, giving 672 weekly positions. For each source, we
counted its records at each position and divided by the number of
observed native intervals at that position. An observed interval with no
records from a particular source contributed zero records and one
interval of exposure. An interval absent from the native reference
contributed neither records nor exposure. The estimated native count for
source $s$ for a target interval $b$ was
\begin{equation}
  \widehat{n}_{s,b} = \frac{C_{s,h(b)}}{m_{h(b)}},
  \label{eq:native-reference}
\end{equation}
where $h(b)$ is its weekday and clock position, $C_{s,h(b)}$ is the
reference record count at that position, and $m_{h(b)}$ is the number of
observed native intervals at that position. All 672 positions had
observations in the pooled reference. Summing across sources gave the
expected count for each target interval; summing across target intervals
gave the expected volume for each source. This estimates recurring
source-specific rates from surrounding weeks and assumes that they apply
during the gap.

For example, the three observed Monday intervals at 12:00 UTC, on 2
June, 9 June, and 7 July, contained 1,291, 1,070, and 1,061 records
across all sources. Their mean, 1,140.7 records, supplied the estimated
native count for each target Monday at 12:00. The estimate describes the
reference's recurring temporal pattern; native records for the missing
Mondays are unavailable.

For the temporal plot, we restricted the reconstruction to records with
minute-level publication-time precision and folded counts into the 672
weekly positions. Each reconstructed count is divided by its number of
target intervals; the native mean divides reference counts by observed
reference exposure. Native weekly rates are scaled to the same sum as
reconstructed weekly rates. The shaded reference range uses the minimum
and maximum of the two or three observed native counts at each position,
with the same scale factor. It describes observed reference variation,
not a confidence interval. Attained volume was calculated separately
using the unscaled native estimate:
\begin{equation}
  A_s = \frac{N_s^{\mathrm{rec}}}{\sum_b\widehat{n}_{s,b}},
  \label{eq:attained-volume}
\end{equation}
where $N_s^{\mathrm{rec}}$ is the eligible output count for source $s$,
and $A_s$ is its attained-volume ratio. The selected-source comparison
uses sources in $S$ in both numerator and denominator; the full
comparison uses every valid native source.

Daily volumes summed all reconstructed records and the unscaled native
estimates over the same target intervals: 24 on 14 June, nine on 2 July,
and 96 on each intervening date. The daily attained ratio divided these
two totals. Daily timestamp-precision bars separated minute-precision
and day-only records. Source-frequency tiers used ranks 1--10, 11--100,
101--1,000, and 1,001 onward in the full pooled native reference.
Tier-by-day ratios divided reconstructed counts by unscaled native
estimates.

\subsection{Archive Discovery and Retrieval}\label{sec:archive-discovery-and-retrieval}
We discovered archived URLs and retrieved their content under a
source-specific allocation. The total proportional allocation was
594,348 articles, corresponding to approximately 30\% of the estimated
native volume for the target period within the selected-source frame
(Section~\ref{sec:reference-model}). Source-specific targets followed
the reference source shares, capped by the available candidate URLs.

We queried the Common Crawl and Internet Archive Wayback Machine indexes
for the selected sources. Source-specific domain and URL-path rules
determined the search scope. For \texttt{iheart.com}, records from the
main site could saturate domain-wide queries and limit coverage of
station subdomains. We therefore queried 704 station hosts separately,
using host names extracted from the reference GKG document URLs.
Discovered records stored the original URL, its normalized key, the
archive capture timestamp, and the information needed to retrieve the
archived page. Repeated captures could refer to the same URL. Archive
capture dates were kept separate from publication dates, which were
assessed during content retrieval.

Before content retrieval, we constructed a URL index, collapsed repeated
normalized URL keys, and preferred records with Common Crawl Web ARChive
(WARC) pointers when available.

Candidates within each source were visited in shuffled order. Archived
content was retrieved through Common Crawl WARC records or Wayback
Machine raw replay URLs using the \texttt{id\_} modifier, and
Trafilatura 2.3.0~\cite{barbaresi2021trafilatura} extracted the main
text and article metadata. Publication dates were obtained from page
metadata, including JSON-LD, HTML metadata, and time elements, with
date-extraction fallbacks. Accepted records required an extracted
publication date within the target period. Publication times were
assigned to 15-minute intervals, with the extraction method and
timestamp precision stored separately.

URL discovery limited the content available for retrieval. Some sources
had no candidates or fewer eligible URLs than their proportional
targets. Download failures and publication-date checks further reduced
the accepted content. The targets defined the intended allocation; the
collected source distribution depended on URL availability and retrieval
success.

The publication-time retrieval window ran from 14 June 2025, 18:00, to 2
July 2025, 02:15 UTC, with the end excluded. The same archive discovery,
content extraction, and eligibility checks applied across the target
interval. Valid retrieved records were retained for preparation; equal
record counts per 15-minute interval were not imposed. For each record,
retrieval provenance identified the archive and capture timestamp.
Common Crawl retrieval used the WARC filename, byte offset, and length;
available locators were retained in the record metadata. Wayback
provenance stored the capture timestamp with the original article URL.
The workflow and its observed counts are shown in
Fig.~\ref{fig:pipeline-funnel}.

\subsection{Text Preparation and Manual Review}\label{sec:text-preparation-and-manual-review}
Before GKG conversion, we screened the extracted bodies for missing
text, binary signatures, JSON payloads, HTML/XML documents,
control-character corruption, and known interface messages. Cleaning
normalized Unicode to NFC, standardized spaces and line breaks, and
removed sparse control characters. Original bodies were preserved for
inspection. Known interface phrases were removed only in a diagnostic
copy to identify boilerplate-only extractions; this test did not strip
passages from accepted bodies. The automated pass rejected 2,489
records: 2,374 boilerplate-only extractions, 68 JSON payloads, 41 markup
payloads, and 6 bodies without alphabetic text.

Bodies with fewer than 80 characters or 40 letters, more than 200,000
characters, less than 20\% alphabetic content among non-whitespace
characters, more than 1\% Unicode replacement characters, or runs of at
least 80 identical characters were flagged for review. These rules
included non-Latin letters and imposed no 50-word minimum. Long
character runs were checked before whitespace normalization, so a
flagged record could contain readable article prose after cleaning.

Language labels were normalized to three-letter source-language codes
(ISO 639-2). For bodies that passed content screening, Lingua
2.2.0~\cite{stahl2026lingua} supplied a language candidate; adoption
required confidence of at least 0.8 and a margin of at least 0.2 over
the next candidate. Proposed disagreements with supplied metadata
underwent pairwise confirmation when supported by the detector. The pass
corrected 2,288 supplied language labels. English bodies were used
directly. Other eligible bodies were translated with M2M100
(Section~\ref{sec:gkg-conversion}).

The automated pass produced 281,001 ready records, 1,545 records
requiring review, and 101 translation failures, in addition to the 2,489
rejections (Table~\ref{tab:preprocessing-outcomes}). Manual inspection
of flagged text accepted 619 records and rejected 919. Seven records in
a language unsupported by the translation model were classified as
translation failures, giving 108 final translation failures and 281,620
eligible records.

\subsection{GKG Conversion}\label{sec:gkg-conversion}
Prepared article bodies and metadata were converted with
\software\ 2.2.0~\cite{klimin2026gdeltregkg} to the 27-column GKG 2.1
layout~\cite{leetaru2015gkgcodebook}. Table~\ref{tab:field-status}
distinguishes constructed identifiers, supplied metadata, approximate
text-derived fields, and unavailable media attributes. Field
definitions, resource identities, and processing configurations are
recorded in the release schema and manifest; reconstruction methods are
described in the processor study~\cite{klimin2026agreement}.

\begin{table}[!h]
\caption{GKG Field Status}
\label{tab:field-status}
\centering\footnotesize
\setlength{\tabcolsep}{3pt}
\begin{tabularx}{\columnwidth}{@{}p{.32\columnwidth}p{.21\columnwidth}X@{}}
\toprule Field group & Status & Construction \\ \midrule
Record ID, date, source, URL & Constructed & Metadata and publication bucket \\
Persons, organizations, locations & Approximate & Entity recognition and geographic resolution \\
Themes & Approximate & Phrase and context rules \\
Tone & Approximate & Lexical scores and densities \\
GCAM & Approximate & Dictionary counts and weighted scores \\
Counts, date mentions, amounts, quotations, all names & Approximate & Extracted values and text spans \\
Translation information, XML extras & Supplied & Language, engine, and page metadata \\
Media (columns 19--22) & Empty & Attributes unavailable \\
\bottomrule
\end{tabularx}
\ArticleNote{Paired agreement for evaluated text fields is reported in
Table~\ref{tab:holdout-agreement}. Complete field encodings are in
\texttt{SCHEMA.md}.}
\end{table}

English bodies were analyzed directly; other eligible bodies were
translated with M2M100
(\texttt{facebook/m2m100\_418M})~\cite{fan2021m2m100}. Named entities
were extracted with spaCy~\cite{honnibal2020spacy} and its
\texttt{en\_core\_web\_lg} model~\cite{explosion2024encoreweblg}.
Geographic resolution, theme rules, lexical tone scoring, and extraction
rules supplied the text-derived fields. GCAM uses an extended profile
with 2,673 configured categories across 34 dictionary IDs. English
dictionaries analyze the English body and language-specific dictionaries
analyze the original body, following the processor
study~\cite{klimin2026agreement}. Enhanced-field offsets refer to the
English analysis text. Extracted publication timestamps determine the
reconstruction's 15-minute buckets; native batch identifiers describe
update time.

\subsection{Timestamp Conventions}\label{sec:timestamp-conventions}
The date audit counted extraction methods and precision labels, checked
the target-window and 15-minute assignments, and compared publication
and archive capture timestamps. Capture lag was capture time minus
stored publication time. For day-only labels, we treated publication as
a 24-hour interval beginning at the stored value and reported bounds on
lag summaries. We audited the first unqualified time element and the
extractor's metadata-date fallback separately. Calendar dates in URL
paths supplied an additional consistency diagnostic; they were not an
independent reference for publication-date accuracy.

Native interval times were taken from the GKG record identifier. Every
usable native \texttt{DATE} value matched that identifier's batch
prefix. We parsed \texttt{PAGE\_PRECISEPUBTIMESTAMP} from XML extras.
GDELT documents this optional field as an extracted UTC publication
timestamp~\cite{gdelt2016articlemetadata}. We measured
batch-to-publication lag and changes in UTC day and weekday among
records with valid tags. Missing publication tags were not imputed. The
temporal reference construction in Section~\ref{sec:reference-model}
continued to use observed batch intervals.

The reconstruction used extracted publication timestamps rounded down to
15 minutes. Comparison of the native and reconstructed profiles
therefore concerns their recorded time distributions; it cannot
establish agreement between native processing time and article
publication time. Records with day-only precision contributed to daily
analyses and volume totals, but were excluded from the primary intraday
plot.

\subsection{Comparison Metrics}\label{sec:comparison-metrics}
We compare aligned normalized distributions with total variation (TV):
\begin{equation}
\operatorname{TV}(p,q)=\tfrac12\sum_i|p_i-q_i|.
\label{eq:total-variation}
\end{equation}
Here, $p_i$ and $q_i$ are the shares of the two distributions in the
same category $i$, with $\sum_i p_i=\sum_i q_i=1$. In comparisons of the
reconstruction with its reference, $p$ denotes the native or modeled
reference distribution and $q$ the reconstructed distribution.
Categories are sources, time bins, or joint source--time cells. For
absent sources, their reference mass gives a lower bound:
\begin{equation}
\operatorname{TV}(p,q)=\delta+\sum_{i:q_i>0}\max(p_i-q_i,0),
\label{eq:tv-decomposition}
\end{equation}
where $\delta=\sum_{i:q_i=0}p_i$. The remaining term is the deficit
among represented sources.

Finite-count benchmarks used 200 independent multinomial draws from each
estimated reference distribution at the observed reconstruction count:
237,465 for minute-precision analyses and 281,620 otherwise.
Source-matched daily draws fixed observed source totals and sampled
dates within each source. Native-window benchmarks sampled 281,620
records without replacement from each window in each of 200 replicates.
These simulations condition on the estimated references rather than
modeling archive selection or uncertainty in native rates.
Appendix~\ref{app:validation-details} reports their means and central
95\% ranges; Appendix~\ref{app:distributions} reports auxiliary
divergences.

Temporal and ordinary joint comparisons used the full native source
frame. Source shares were calculated separately within the full
reference frame and the selected-source frame. Sources with no output
retained zero shares. For joint comparisons, we summed the expected
source/interval counts into source-by-target-day and
source-by-weekday/quarter tables, then normalized expected and observed
tables to their respective totals. A further daily comparison rescaled
each expected source total to its observed total before normalization,
isolating within-source temporal differences. Absent sources contributed
to the ordinary joint comparison but had no observed conditional
temporal profile.

\subsection{Content Comparison}\label{sec:content-comparison}
We compared the complete reconstruction with native records from 7--13
June and 2--8 July, using observed batches on the partially covered date
of 2 July. The first window supplies seven days before the outage
without including the interrupted date of 14 June. For each article, we
counted each theme code and each country code in the standard location
field once, regardless of repeated mentions. We divided these counts by
the article total to obtain label prevalence, and normalized the
prevalence vectors across labels for TV. We reported the fractions of
articles containing any theme or location separately. Country codes
describe places mentioned in text, not publisher countries. Tone
comparisons used the first component of \texttt{V2Tone}, its article
mean, and fixed two-point bins spanning $[-100,100]$.

Each comparison used both complete populations and a common-source
panel. On the panel, articles from source $s$ in corpus $c$ received
weight $N_s^{\mathrm{rec}}/N_{s,c}$, so both corpora had the same
reconstructed source totals before normalization. Sources absent from
either corpus were excluded from that comparison, and their support was
reported. The same procedure compared the two native weeks, providing a
descriptive reference for adjacent-period differences. This content
audit reads generated GKG fields without rerunning NLP.

The daily illustration uses native records with parsed publication tags
no later than batch time and reconstructed articles with English input.
It pools articles by extracted UTC publication date. Its standardized
curves weight publisher means by fixed pooled native source shares on
sources represented before, during, and after the outage, renormalized
over sources observed each day. This illustration uses native source
weights; the population comparisons above use reconstructed source
weights.

\subsection{Intact-Week Reconstruction}\label{sec:intact-week-reconstruction}
We tested the reconstruction process on 2--8 June 2025 to measure
archive recovery and field agreement against available native GKG
records. All 672 native batches are present for this week. Articles were
recovered from Common Crawl and the Wayback Machine using the same
source-guided discovery, retrieval, text preparation, review, and GKG
conversion workflow as the outage dataset. Archive captures were
searched through 15 July; extracted publication dates were restricted to
the test week.

Unlike the outage reconstruction, this experiment used source
frequencies from 19 May--1 June and 9--15 June, excluding the test week
from source selection and quota estimation. The same eligibility rules
selected 5,530 sources. URLs and content fields from original GKG files
were reserved for evaluation and did not guide archive discovery.
Comparisons used the full eligible collection and a source-balanced
subset of 5,000 articles; the matching and scoring procedures are
described in Appendix~\ref{app:holdout}.

\section{Data Records}\label{sec:data-records}
Version 1.0.0 is available on Zenodo~\cite{klimin2026regkgdataset} as a
12-file package of approximately 1.16~GiB
(Table~\ref{tab:release-files}). The outer \texttt{data.zip} contains
\texttt{data/gkg/YYYYMMDDHHMMSS.gkg.csv.zip} and the master file list.
Each batch ZIP contains one UTF-8, tab-separated, headerless file with
27 GKG columns. The outer archive stores the already compressed batch
ZIPs without recompressing them, preserving their original bytes and
record identifiers. Master-list entries give batch ZIP byte size, MD5,
and relative path; the manifest provides SHA-256 identities for both
archives and uncompressed members.

\begin{table}[!t]
\caption{Version 1.0.0 Release Files}
\label{tab:release-files}
\centering\footnotesize
\setlength{\tabcolsep}{3pt}
\newcommand{\ReleaseFile}[3]{\texttt{#1} & #2 \\
  & #3\\}
\begin{tabularx}{\columnwidth}{@{}>{\raggedright\arraybackslash}X r@{}}
\toprule File & Contents and size \\ \midrule
\ReleaseFile{data.zip}{GKG batch archives and master list}{1.10~GiB}
\ReleaseFile{record\_provenance.parquet}{Record-level provenance}{22.61~MiB}
\ReleaseFile{archive\_provenance.parquet}{Archive replay/range locators}{31.35~MiB}
\ReleaseFile{source\_day\_weights.parquet}{Source/day calibration weights}{1.11~MiB}
\ReleaseFile{review\_journal.parquet}{Manual review decisions}{122.96~KiB}
\ReleaseFile{sources.csv}{Source allocation and attained volume}{1.37~MiB}
\ReleaseFile{README.md}{Dataset description, validation, and use}{13.87~KiB}
\ReleaseFile{SCHEMA.md}{GKG and sidecar definitions}{14.88~KiB}
\ReleaseFile{LICENSE.txt}{CC BY 4.0 legal text}{16.62~KiB}
\ReleaseFile{NOTICE.md}{Attribution and third-party rights}{1.65~KiB}
\ReleaseFile{manifest.json}{Inventory and embedded profiles}{865.27~KiB}
\ReleaseFile{checksums.sha256}{File checksums}{917~B}
\midrule
Total (12 files) & 1.16~GiB \\
\bottomrule\end{tabularx}
\end{table}

The record provenance table has one row per released record, keyed by
\texttt{record\_id}. It includes URL, source, archive, capture and
publication timestamps, assigned batch, date precision and method,
language, translation flag, and manual acceptance flag. Its
\texttt{archive\_id} joins to the archive provenance table. Every Common
Crawl entry has a WARC filename and compressed byte range. Wayback
entries contain the capture label and replay address.

The \texttt{sources.csv} table contains all 12,459 native source
identities, including the 5,548 selected sources and zero-output
sources. It reports quotas, candidates, final eligibility counts,
reference shares, estimated native volume, attained fraction, and quota
fill. The source/day Parquet has 236,721 cells, with expansion and
composition weights on supported cells.

The full schema and precision conventions are documented in
\texttt{SCHEMA.md}. Columns 19--22 are empty in every record:
\texttt{V2.1SHARINGIMAGE}, \texttt{V2.1RELATEDIMAGES},
\texttt{V2.1SOCIALIMAGEEMBEDS}, and \texttt{V2.1SOCIALVIDEOEMBEDS}. The
corresponding image and social-media attributes were not collected. An
empty supported text field means that the calculation produced no
matches.

The file \texttt{manifest.json} embeds the definitions referenced by all
processing and translation profile links. These definitions identify the
field resources, translation configuration, and recorded calculations.

\section{Technical Validation}\label{sec:validation}
We ask five questions about the release. The first four use the full
release dataset. The fifth uses the intact-week test
(Section~\ref{sec:intact-week-reconstruction}), where native
counterparts exist. Native weeks outside the gap serve as references.
Because they cover different events, we compare each
release-to-reference distance with the distance between two native weeks
and with random sampling noise.

\subsection{Are the Articles Valid?}\label{sec:are-the-articles-valid}
Of 285,136 retrieved records, 281,620 (98.8\%) were eligible. Automatic
screening rejected 2,489 records, manual review of 1,545 flagged records
rejected 919, and 108 failed translation
(Fig.~\ref{fig:pipeline-funnel},
Table~\ref{tab:preprocessing-outcomes}). Most rejections were
boilerplate-only pages. The eligible set contains 256,617 Wayback and
25,003 Common Crawl records, of which 273,108 are English and 8,512
(3.0\%) translated. Every record has a unique GKG identifier and
source--URL pair, batch prefixes match the assigned buckets, per-batch
counts match the manifest, all 1,665 intervals are nonempty, and all
checksums verify. Manual review covered only flagged records, so the
share of invalid articles among the accepted ones is not estimated
(Section~\ref{sec:limitations}).

\subsection{Can the Timestamps be Trusted?}\label{sec:can-the-timestamps-be-trusted}
\begin{figure}[!htbp]
\centering
\includegraphics[width=\columnwidth]{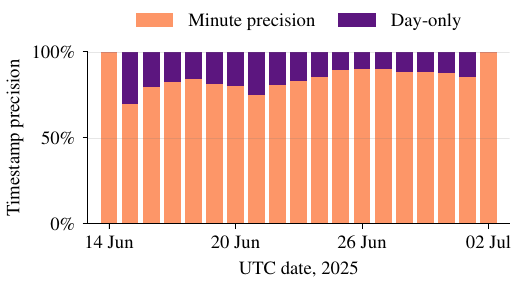}
\caption{Daily publication-time precision. Shares use daily reconstructed totals. The first and last dates have partial exposure.}
\label{fig:timestamp-precision}
\end{figure}

Publication dates come from page metadata. Extracted timestamps have
minute-level precision for 84.3\% of records. The remaining 15.7\% have
day-only dates (Fig.~\ref{fig:timestamp-precision}). Most of these dates
come from the extractor's fallback (Table~\ref{tab:date-extraction}),
and 38,568 of the 44,155 day-only records fall into the 00:00 UTC
bucket. This creates a midnight spike and raises the temporal TV from
0.171 to 0.244 (Section~\ref{sec:how-much-was-recovered}). Daily
analyses can use all records; intraday analyses should use the
minute-precision subset. The median archive capture lag among
minute-precision records is 5.7 hours for Wayback and 1.2 hours for
Common Crawl. In 8,062 records (2.9\%) capture precedes the recorded
publication time. These are flagged in the provenance table and may
reflect extraction errors or page changes.

\begin{figure}[!htb]
\centering
\includegraphics[width=\columnwidth]{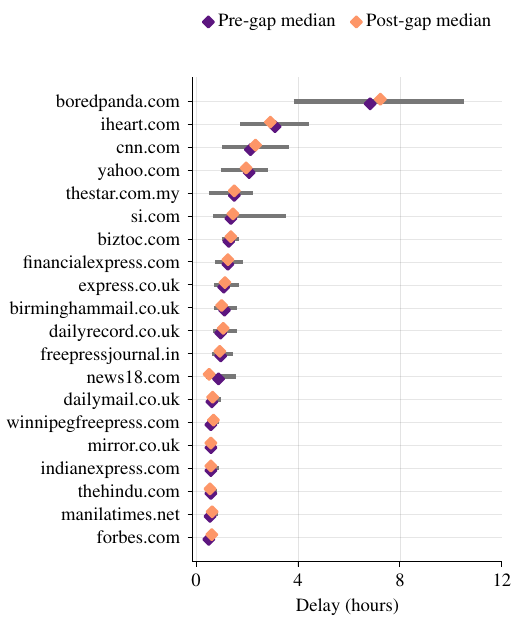}
\caption{Native publication-to-batch delay for the 20 most frequent
supported sources with pooled upper quartile no greater than 12 hours.
Gray bars show pooled within-source IQRs.}
\label{fig:source-delay-intervals}
\end{figure}

\begin{figure}[!htbp]
\centering
\includegraphics[width=\columnwidth]{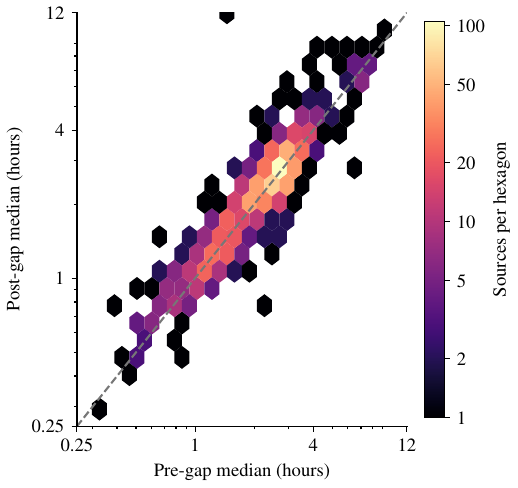}
\caption{Pre-gap versus post-gap median publication-to-batch delay
by source. Hexagons count sources; color shows
source counts on a logarithmic scale. The gray line denotes equality.
The plot excludes 19 sources with medians outside
0.25--12 hours.}
\label{fig:source-delay-hexbin}
\end{figure}

Native batch time is not publication time. For the 58.7\% of native
records with a publication tag, batch time lags publication by a median
of 129 min (IQR 57--240 min). Among sources with at least 100
nonnegative tagged lags in each period, pre-gap and post-gap source
medians have Spearman correlation of 0.90 and a median absolute change
of 13 minutes (Figs.~\ref{fig:source-delay-intervals}
and~\ref{fig:source-delay-hexbin}).

Within the supported sources, the median within-source IQR is 143 min,
so a fixed shift cannot reproduce the observed variation in native batch
times. ReGKG buckets use extracted publication timestamps without a
delay adjustment; they do not show when a record would have been
available in real time. Appendix~\ref{app:processing-delay} examines
optional delay scenarios and their boundary effects.

\subsection{How Much was Recovered?}\label{sec:how-much-was-recovered}
\begin{figure}[!htbp]
\centering
\includegraphics[width=\columnwidth]{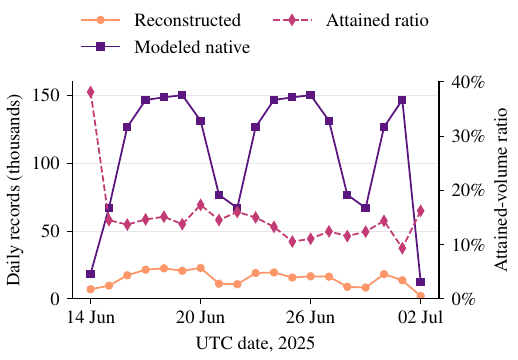}
\caption{Daily reconstructed and modeled native counts (solid lines, left axis), and their ratio (dashed line, right axis).}
\label{fig:daily-volume}
\end{figure}

Applying source-specific weekday and time-of-day rates from the
surrounding weeks (Section~\ref{sec:reference-model}), the gap would
contain about 2.07 million records across all sources, or 1.97 million
in the selected sources. The release holds 13.6\% and 14.3\% of these,
respectively. Daily ratios range from 9\% to 17\%
(Fig.~\ref{fig:daily-volume}), with the exception of 14 June at 38\%.

The 14 June estimate covers only 18:00--24:00 UTC. All 7,047 records
have minute precision, and Common Crawl supplies 59.5\% of them,
compared with 8.9\% overall, giving this boundary date a different
retrieval mix. Its attained ratio remains higher when other dates are
restricted to minute-precision records (8.0--16.2\%). We have not
identified the cause and recommend treating this date separately in
daily-ratio analyses.

\begin{figure}[!htbp]
\centering
\includegraphics[width=\columnwidth]{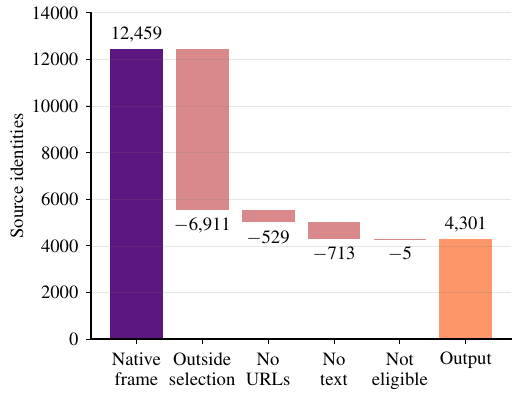}
\caption{Source coverage. Waterfall losses use mutually exclusive sets defined by source selection, candidate availability, retrieved text, and eligible output.}
\label{fig:source-losses}
\end{figure}

The release contains 4,301 of the 12,459 native sources. Of the 8,158
absent ones, 6,911 did not meet the selection criteria, 529 had no
archived URLs, 713 had URLs but no retrievable text, and 5 had no
eligible articles (Fig.~\ref{fig:source-losses}). The 1,247 selected
sources that are missing hold 16.1\% of the selected reference mass, so
the release covers 83.9\% of it. Coverage depends on source frequency.
Eight of the ten most frequent sources are present. Sources ranked
11--100 are the weakest tier, with 66\% present and 8.7\% of output
against 13.1\% of native mass. Sources ranked 101--1,000 are best
covered (88\% present). The full tier breakdown is in
Table~\ref{tab:frequency-tiers} in Appendix~\ref{app:distributions}.

\begin{table}[!t]
  \caption{Distance to Reference}
  \label{tab:distance-to-reference}
  \centering
  \footnotesize
  \setlength{\tabcolsep}{3pt}
  \begin{tabularx}{\columnwidth}{@{}Xrrr@{}}
    \toprule
    Comparison & Release & Random & Native \\
    \midrule
    Source shares, all sources & 0.382 & 0.054 & 0.127 \\
    Source shares, selected sources & 0.358 & 0.045 & 0.100 \\
    Weekly profile (minute precision) & 0.171 & 0.033 & --- \\
    Weekly profile (all) & 0.244 & 0.030 & --- \\
    Source $\times$ day & 0.503 & 0.206 & --- \\
    Source $\times$ day, source totals matched & 0.292 & 0.144 & --- \\
    \bottomrule
  \end{tabularx}
  \ArticleNote{All values are TV. Release compares the release with the reference. Random reports mean simulated TV (Section~\ref{sec:comparison-metrics}). Native compares equal-size samples from the two native windows.}
\end{table}

\begin{figure}[!htbp]
\centering
\includegraphics[width=\columnwidth]{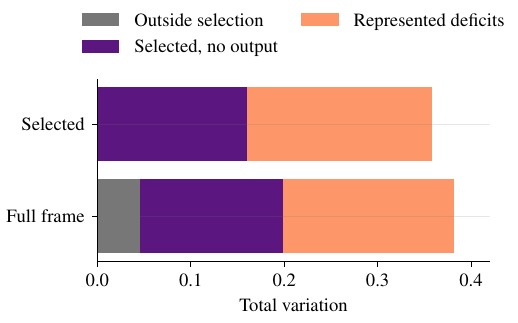}
\caption{Source-share TV decomposed into missing reference mass and remaining deficits among represented sources. The components sum to TV, including the compensating reallocation of absent-source mass.}
\label{fig:source-tv-decomposition}
\end{figure}

Table~\ref{tab:distance-to-reference} compares the release with the
reference, with sampling benchmarks, and with equal-count samples from
the native pre-gap and post-gap windows. The source mix differs from the
reference by TV 0.36 (0.38 over all sources), three to four times the
equal-count native-window distances (0.10--0.13) and about eight times
the selected-source sampling benchmark. Missing selected sources account
for 0.161 of selected-frame TV (Fig.~\ref{fig:source-tv-decomposition}).
The remaining term is the reference-share deficit among represented
sources. The largest individual source-share difference is for
\texttt{iheart.com}: 8.6\% of output against 6.0\% of the full native
reference (Table~\ref{tab:largest-source-errors}).

\begin{figure}[!h]
\centering
\includegraphics[width=\columnwidth]{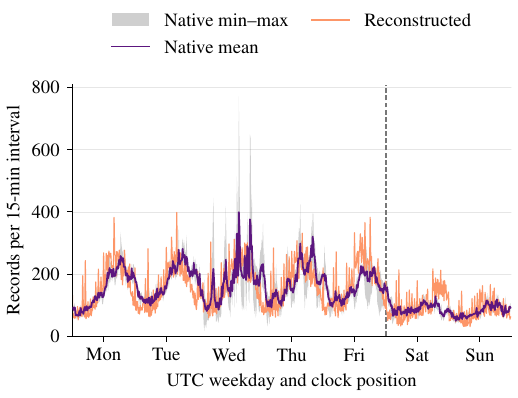}
\caption{Folded weekly profile on 672 weekday/quarter-hour positions. Reconstructed minute-precision counts are divided by the number of target intervals at each position. The native mean and observed min--max use actual observed reference exposure and are scaled to the same sum of weekly rates. The vertical dashed line marks the start of the weekend. This is a recurring weekly profile, not native observations during the outage; the band is not a confidence interval.}
\label{fig:weekly-profile}
\end{figure}

The broad weekly pattern is similar (Fig.~\ref{fig:weekly-profile}),
with TV 0.171 for minute-precision records against 0.033 for noise.
Jointly by source and day, TV is 0.503 and falls to 0.292 when each
source's total is matched to its expected total. Source-total matching
reduces the joint distance, but within-source temporal differences
persist. The finer source-by-weekday/quarter grid is too sparse for a
stable comparison (Appendix~\ref{app:distributions}).

\subsection{Does the Content Match Native Data?}\label{sec:does-the-content-match-native-data}
We compared the release with native records from 7--13 June (839,881)
and 2--8 July (817,543). Release-to-native differences exceed the
observed differences between the two native weeks
(Table~\ref{tab:content-comparison}). Theme TV is 0.275--0.278 against
0.075 between the native weeks. Mentioned-country TV is 0.126--0.149
against 0.078, and tone TV is 0.105--0.108 against 0.012. The
differences persist after standardizing source composition on common
support.

\begin{table}[H]
  \caption{Adjacent-Week Content Distances}
  \label{tab:content-comparison}
  \centering
  \footnotesize
  \setlength{\tabcolsep}{3pt}
  \begin{tabular*}{\columnwidth}{@{\extracolsep{\fill}}lrrr@{}}
    \toprule
    Comparison & Themes & Countries & Tone \\
    \midrule
    \multicolumn{4}{l}{\textit{Complete populations}} \\
    Pre / gap & 0.275 & 0.126 & 0.105 \\
    Post / gap & 0.278 & 0.149 & 0.108 \\
    Pre / post & 0.075 & 0.078 & 0.012 \\
    \midrule
    \multicolumn{4}{l}{\textit{Common sources, gap source weights}} \\
    Pre / gap & 0.268 & 0.126 & 0.103 \\
    Post / gap & 0.272 & 0.131 & 0.107 \\
    Pre / post & 0.082 & 0.069 & 0.017 \\
    \bottomrule
  \end{tabular*}
  \ArticleNote{Entries are TV distances. Countries denotes mentioned-country codes. Comparison windows and source standardization are defined in Section~\ref{sec:content-comparison}.}
\end{table}

Mean tone is $-0.64$ in the release and $-0.68$ and $-0.66$ in the two
native weeks. On common-source panels with native records weighted to
the release's source totals, native mean tone is about $-0.95$, while
the reconstructed mean is about $-0.64$. Under this standardization, the
release is about 0.3 points less negative. The apparent agreement of
pooled means therefore does not establish content comparability.
Analyses spanning the outage should assess possible discontinuities in
tone and theme shares at the change in observation regime.

\begin{figure}[!h]
\centering
\includegraphics[width=\columnwidth]{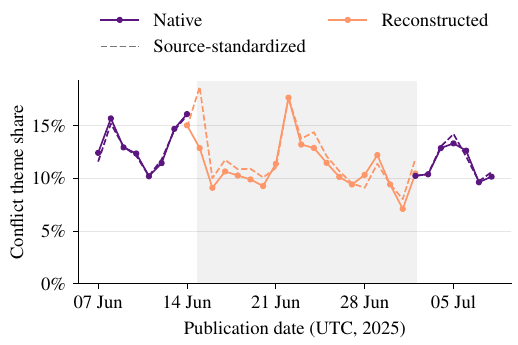}
\caption{Daily prevalence of the ARMEDCONFLICT theme. Solid curves pool
articles; dashed curves standardize publisher means to fixed native
source shares on common support, renormalized daily. Shading marks the
outage; boundary dates have partial exposure.}
\label{fig:daily-content-example}
\end{figure}

Fig.~\ref{fig:daily-content-example} shows the daily share of articles
with the ARMEDCONFLICT theme on the publication-time subsets described
in Section~\ref{sec:content-comparison}. The raw reconstructed shares
range from 7.1\% to 17.7\%, overlapping the native range of 9.6--16.1\%
but extending beyond it. The partial-day values at both boundaries are
close, without an obvious step in this plot. The raw reconstructed
series peaks on 22 June at 17.7\%, above the observed native range.

Elevated conflict-theme shares around 13--15, 22 and 24 June coincided
with military escalation and ceasefire developments in the Middle
East~\cite{unsc2025middleeast9936,unsc2025middleeast9941,unsc2025middleeast9944}.
The rise around 29 June coincided with intensified hostilities in
Europe~\cite{unsc2025europe9965}, however its magnitude remained within
the observed native day-to-day range, so it does not establish an
event-specific response. These correspondences provide historical
context without attributing the observed peaks to specific events.

Source standardization changes shares by less than one percentage point
on most dates, but raises 15 June from 12.9\% to 18.7\%. The lack of an
obvious boundary step for this theme does not establish continuity for
other themes, given the large overall theme distance in
Table~\ref{tab:content-comparison}; continuity should be checked per
theme.

\newsavebox{\ArticleHoldoutCoverage}
\begin{lrbox}{\ArticleHoldoutCoverage}
\begin{minipage}{\columnwidth}
\begin{table}[H]
\caption{Intact-Week Archive Recovery}\label{tab:holdout-coverage}
\centering\footnotesize\setlength{\tabcolsep}{3pt}
\begin{tabular*}{\columnwidth}{@{\extracolsep{\fill}}lrr@{}}
\toprule Measure & Full pool & Balanced \\ \midrule
Prepared articles & 14,437 & 5,000 \\
Represented sources & 3,132 & 3,132 \\
Publication buckets & 672 & 648 \\
Exact source--URL matches & 5,918 & 2,081 \\
Match rate (\%) & 40.99 & 41.62 \\
URL recall (\%) & 0.700 & 0.246 \\
Source TV: reference target & 0.543 & 0.343 \\
Source TV: native week & 0.552 & 0.370 \\
\bottomrule\end{tabular*}
\ArticleNote{Native frame: 845,948 unique source--URL pairs. Match rate uses recovered articles as its denominator; URL recall uses the native frame. Subset construction is described in Appendix~\ref{app:holdout}.}\end{table}

\end{minipage}
\end{lrbox}
\begin{table*}[!b]
\caption{Paired Agreement in the Intact Week and the Processor Study}\label{tab:holdout-agreement}
\centering\footnotesize\setlength{\tabcolsep}{3pt}
\begin{tabular*}{\textwidth}{@{\extracolsep{\fill}}lrrrrrrr@{}}
\toprule Population & Pairs & Tone MAE & Tone CCC & Persons & Organizations & Locations & Themes \\ \midrule
Processor study: English & 193,550 & 1.790 & 0.749 & 0.646 & 0.417 & 0.654 & 0.617 \\
Intact week: all pairs & 5,918 & 1.690 & 0.806 & 0.673 & 0.456 & 0.648 & 0.638 \\
Intact week: English source & 5,891 & 1.689 & 0.807 & 0.674 & 0.458 & 0.650 & 0.639 \\
Intact week: balanced subset & 2,081 & 1.697 & 0.799 & 0.674 & 0.441 & 0.650 & 0.634 \\
Processor study: title/length & 54,485 & 1.470 & 0.830 & 0.748 & 0.483 & 0.727 & --- \\
Intact week: title/length & 1,848 & 1.470 & 0.851 & 0.725 & 0.506 & 0.686 & 0.671 \\
\bottomrule\end{tabular*}
\ArticleNote{Categorical values are micro F1; themes use all observed labels. Tone MAE is in GKG points. Title/length subsets use a 10\% length band but different title eligibility and normalization. A dash means the score was not reported in the cited table.}\end{table*}

\begin{table*}[!b]
\caption{Tasks and Conditions of Use}\label{tab:usage}
\centering\footnotesize
\begin{tabularx}{\textwidth}{@{}p{.23\textwidth}p{.15\textwidth}X@{}}
\toprule Task & Suitability & Conditions \\ \midrule
Retrospective content series & Supported with controls & Report sample size, source composition, timestamp precision, and regime boundaries. \\
Source-level daily comparison & Conditional & Restrict to supported cells; report weights and missing-source mass. \\
GKG-format pipeline testing & Supported & Read all 27 fields and distinguish four unavailable media columns from no text matches. \\
Gap-handling benchmarks & Conditional & Evaluate against observed intact periods; preserve archive selection and document-version uncertainty. \\
Exhaustive article/event counts & Unsupported & Modeled-volume calibration does not recover absent publications or identify inclusion probabilities. \\
Real-time trading or monitoring & Unverified & Publication and archive capture times do not establish historical first availability. \\
\bottomrule\end{tabularx}
\end{table*}

\subsection{Intact-Week Reconstruction and Test}\label{sec:how-well-do-the-generated-fields-agree}
The processor study measured field agreement on 193,550 English
articles~\cite{klimin2026agreement}. We repeated agreement assessment in
the intact-week test (Section~\ref{sec:intact-week-reconstruction}).
Articles for 2--8 June were recovered with the same pipeline, and 5,918
of the 14,437 eligible articles (match rate 41.0\%) matched native
records by exact source and URL (Table~\ref{tab:holdout-coverage}).
Native GKG is a sample, not a census, so absence from the native set
does not establish an acquisition error. The pool is 1.7\% of the native
week's volume; its 41.0\% match rate gives URL recall of 0.7\%. On
matched pairs, tone MAE is 1.69 (CCC 0.81), and micro F1 is 0.67 for
persons, 0.46 for organizations, 0.65 for locations and 0.64 for themes
(Table~\ref{tab:holdout-agreement}). Tone MAE and location F1 are close
to the processor-study values of 1.79 and 0.65, respectively. These are
descriptive comparisons across different article populations; the
generated fields are approximations rather than verified annotations.

At the source level, the quota reference has TV of 0.074 to the native
week. The recovered pool has TV of 0.54 to the reference target, or 0.34
after source balancing, indicating that the recovered source mixture
differs substantially from the intended allocation. The pool's mean tone
is $+0.14$ against $-0.63$ in the native week
(Table~\ref{tab:holdout-content}). On common sources, weighting native
records to recovered source totals leaves a difference of $+0.70$
points, larger than the release's approximately $+0.3$. Both exceed the
$+0.11$ mean paired bias, so paired agreement alone does not establish
comparability of population-level content measures. The size of the tone
shift therefore differs between collections, so the intact week cannot
be used to correct the release, and the shift is not explained by
tone-coding error alone.

\noindent\usebox{\ArticleHoldoutCoverage}\par

\section{Usage Notes}\label{sec:usage-notes}\label{sec:discussion}

Table~\ref{tab:usage} relates common tasks to the dataset's observation
scope. Analyses crossing the outage should retain a native/reconstructed
indicator and report article counts used as denominators, the sources
represented in each aggregate, and timestamp precision. Daily aggregates
can include day-only records. Intraday analyses should use the
minute-precision subset and distinguish publication time from assumed
processing or availability time. Reconstructed batch labels represent
publication timestamps rounded down to 15-minute intervals, whereas
native batch labels represent update times. Analyses combining the two
should specify a common time basis.

The following example reads one GKG batch directly from the release
archive and joins its record provenance on \texttt{record\_id}, without
extracting the whole release. It checks that every batch record has a
provenance match. Archive provenance can then be joined separately on
\texttt{archive\_id}.
\begin{lstlisting}[language=Python,basicstyle=\ttfamily\scriptsize,breaklines=true]
from io import BytesIO
from zipfile import ZipFile
import csv
import pandas as pd

root = "regkg-2025-v1.0.0/"
with ZipFile(root + "data.zip") as outer:
  name = "data/gkg/20250614180000.gkg.csv.zip"
  with ZipFile(BytesIO(outer.read(name))) as batch:
    with batch.open(batch.namelist()[0]) as stream:
      records = pd.read_csv(stream, sep="\t", header=None,
        dtype=str, keep_default_na=False,
        quoting=csv.QUOTE_NONE)
records = records.rename(columns={0: "record_id"})
provenance = pd.read_parquet(
  root + "record_provenance.parquet")
joined = records.merge(
  provenance, on="record_id", how="left",
  validate="one_to_one", indicator=True)
assert joined["_merge"].eq("both").all()
joined = joined.drop(columns="_merge")
\end{lstlisting}

Recovery rates differ across sources and days, so unweighted aggregates
can give a different source mixture from the modeled native reference.
The file \texttt{source\_day\_weights.parquet} provides calibration
weights for adjusting source contributions. Its expansion factor
calibrates observed article counts to modeled native volume:
\begin{equation}
  w^{\mathrm{exp}}_{sd}=e_{sd}/n_{sd},\qquad n_{sd}>0,\ e_{sd}>0.
\end{equation}
Here $s$ denotes the source, $d$ the day, $n_{sd}$ the reconstructed
article count, and $e_{sd}$ the modeled native count. Cells with both
counts positive form the supported daily panel. The file also provides a
composition factor that adjusts source shares while preserving the
observed article total on that panel:
\begin{equation}
  w^{\mathrm{comp}}_{sd}=
    \frac{e_{sd}/E_d}{n_{sd}/N_d},
\end{equation}
where $E_d$ and $N_d$ sum expected and observed counts only over cells
with both quantities positive. Weighted observed counts sum to $E_d$
under expansion and $N_d$ under composition.

To calculate a daily weighted mean, assign each article the weight of
its source/day cell, sum the products of article values and weights, and
divide by the sum of weights. Article values can be tone scores or
binary indicators of a theme label; the latter yield weighted label
shares. Both factors give the same daily weighted mean on the same
supported cells because they differ only by the daily constant
$N_d/E_d$.

Undefined factors are null; they must not be filled with zero or
infinity. Report excluded cells, reference mass, and weight dispersion.
These calibration factors are not inverse-probability sampling weights
because article inclusion probabilities are unknown.

For field-based analyses, the intact-week benchmark provides direct
agreement estimates on recovered documents
(Table~\ref{tab:holdout-agreement}). Person, organization, location and
theme labels have different precision and recall. Use explicit field
definitions and sensitivity analyses rather than treating a
reconstructed label as a verified annotation. Source balancing reduces
publisher mixture differences but does not calibrate field errors or
selection within a publisher.

\section{Limitations and Ethics}\label{sec:limitations-and-ethics}
\subsection{Limitations}\label{sec:limitations}
The reconstruction is an accessible-archive sample with unequal recovery
across publishers and articles. Its 13.63--14.28\% attained-volume
ratios use surrounding-week estimates, not observed missing-period
totals or known inclusion probabilities. Sources absent from the output
account for 16.06\% of selected reference mass. Reweighting represented
sources cannot restore those sources or correct unknown selection within
a publisher. Exhaustive article or event counts are outside this scope.

The reference contains two or three observations per weekly clock
position and concerns different news events from the missing period.
Sparse source-by-weekday/quarter cells contribute to the joint distances
(Appendix~\ref{app:distributions}), and absence from those reference
cells is not evidence that a source could not publish then. Content
differences across adjacent weeks combine changing events, article
selection, and differences in coding. Source standardization addresses
publisher mixture on common support; it does not isolate the other
components or provide missing-period ground truth. The native reference
is the English processing stream, whereas 3.02\% of reconstructed
records have non-English source labels.

Publication timestamps, native batch labels, archive capture times, and
object upload times measure different stages. Optional native
publication tags cover only 58.66\% of records and include negative and
extreme lags. The latency analysis therefore describes the tagged
subset; historical first-availability logs would be needed to measure
bucket upload latency. A fixed offset does not reproduce heterogeneous
source-specific delays. Day-only timestamps do not support an hourly
shift. Captures can postdate publication by hours or days, so
retrospective reconstruction and assumed delay scenarios do not
establish that features were available in real time.

The 8,062 capture/publication conflicts could arise from date
extraction, page-version changes, or provenance. URL-date checks can
reuse the extractor's evidence or refer to another event and therefore
are not independent date-error estimates. Midnight values in native
publication tags have no precision label. An independent date audit is
needed before treating these timestamps as verified clock times.

Manual review covered the 1,545 flagged records rather than a
probability sample of all accepted text. Language and translation
accuracy were not independently estimated. The processor's paired
field-agreement study concerns other articles and reports substantial
differences by field; its scores are not dataset-specific accuracy
estimates. An independently annotated probability sample would address
article eligibility and timestamp accuracy.

The intact-week experiment supplies paired agreement on recovered
documents, conditional on archive accessibility and preparation. It uses
one publication week and an archive capture horizon through 15 July,
with acquisition constrained by source-specific budgets. Its
source-balanced subset changes the observed mixture by construction and
has source TV of 0.343 to the reference target. The full pool is
approximately 1.71\% of the observed native week, whereas the outage
release reaches approximately 14\% of modeled volume. The experiment
does not estimate recall inside the actual outage or establish that its
paired-disagreement distribution transfers to the release. Cross-study
agreement comparisons involve different populations and
title-normalization criteria; no significance or equivalence test is
implied. Exact URL matches and matching titles do not prove that the
archived and native article bodies are identical.

\subsection{Ethics and licensing}\label{sec:ethics-legal}
Article collection used existing captures from Common Crawl and the
Internet Archive Wayback Machine. Direct scraping of publisher websites
was not used for collection. URL discovery used the archives' documented
index interfaces~\cite{commoncrawlfaq,internetarchivecdxapi}; content
retrieval used Common Crawl WARC records and Wayback replay URLs. The
collection clients implemented separate request pacing and concurrency
controls for the archive services, with bounded retries and backoff
after rate-limiting and server-error responses.

Full article bodies are not included in the release. The release scope
comprises reconstructed GKG records and provenance metadata. Archive
access does not establish permission to redistribute the underlying
reporting: Common Crawl's terms require respect for third-party rights
and recognize that archived content may have separate terms of
use~\cite{commoncrawl2024terms}. Software and dictionary licenses apply
to their respective components and do not license the source articles.
The creator-held dataset and documentation rights use CC BY 4.0;
third-party rights and attribution are stated in \texttt{NOTICE.md}.

The reconstruction preserves personal names in entity fields and can
include passages from the articles in quotation fields. The pipeline
does not anonymize personal names, so the derived records can identify
people mentioned in the reporting. Names and quotations are extracted
automatically and may contain errors. They record textual mentions and
do not independently verify a person's identity, role, or the claims
made in an article.

\section{Conclusion}\label{sec:conclusion}
ReGKG-2025 covers all 1,665 missing GKG batches with 281,620 article
records, reaching approximately 14\% of modeled native volume. Its
content distributions differ from native GKG, so analyses spanning the
outage must account for the change between native and reconstructed
observations.

\section*{Data and Code Availability}
ReGKG-2025 version 1.0.0 is available on Zenodo under CC BY 4.0 at
\url{https://doi.org/10.5281/zenodo.23964265}~\cite{klimin2026regkgdataset}.
Dataset mirrors are hosted on Hugging
Face\footnote{\url{https://huggingface.co/datasets/naklimin/ReGKG-2025}}
and
Kaggle\footnote{\url{https://www.kaggle.com/datasets/nikitaaklimin/regkg-2025}}.
The package includes a manifest, checksums, a dataset README, and
complete record and sidecar schemas. The gdelt-regkg processing software
is available at PyPI\footnote{\pypi} and GitHub\footnote{\github}.
Analysis scripts are not publicly released. The version 1.0.0 package
does not contain intact-week experimental records.

\appendices
\setcounter{table}{0}
\counterwithin*{table}{section}
\renewcommand{\thetable}{\Alph{section}\arabic{table}}
\renewcommand{\theHtable}{appendix.\Alph{section}.\arabic{table}}
\setcounter{figure}{0}
\counterwithin*{figure}{section}
\renewcommand{\thefigure}{\Alph{section}\arabic{figure}}
\renewcommand{\theHfigure}{appendix.\Alph{section}.\arabic{figure}}
\section{Validation Details}\label{app:validation-details}
Table~\ref{tab:preprocessing-outcomes} accounts for all 285,136
retrieved records across automated preparation, manual review, and final
conversion eligibility. The final input combines automatically ready
records with accepted reviewed records; unsupported-language cases are
counted as translation failures.

\begin{table}[!th]
  \caption{Preprocessing Outcomes}
  \label{tab:preprocessing-outcomes}
  \centering
  \footnotesize
  \setlength{\tabcolsep}{3pt}
  \begin{tabular*}{\columnwidth}{@{\extracolsep{\fill}}lr@{}}
    \toprule
    Outcome & Records \\
    \midrule
    \multicolumn{2}{l}{\textit{Automated Preparation}} \\
    Ready & 281,001 \\
    Flagged for review & 1,545 \\
    Rejected & 2,489 \\
    Translation failed & 101 \\
    \midrule
    \multicolumn{2}{l}{\textit{Manual Review}} \\
    Accepted & 619 \\
    Rejected & 919 \\
    Unsupported translation language & 7 \\
    \midrule
    \multicolumn{2}{l}{\textit{Final Conversion Input}} \\
    Ready & 281,620 \\
    Rejected & 3,408 \\
    Translation failed & 108 \\
    \bottomrule
  \end{tabular*}
\end{table}

Table~\ref{tab:sampling-benchmarks} gives the empirical means and
central 95\% ranges from 200 replicates per comparison, described in
Section~\ref{sec:comparison-metrics}. The reference-draw benchmarks
condition on the estimated probabilities; native-window subsampling
conditions on the two observed collections. Neither range estimates
uncertainty in the unavailable gap-period distribution.

\begin{table}[H]
  \caption{Sampling Benchmarks}
  \label{tab:sampling-benchmarks}
  \centering
  \footnotesize
  \setlength{\tabcolsep}{3pt}
  \begin{tabular*}{\columnwidth}{@{\extracolsep{\fill}}lrrrr@{}}
    \toprule
    Comparison & $N$ & Mean TV & 2.5th & 97.5th \\
    \midrule
    \multicolumn{5}{l}{\textit{Independent Reference Draws}} \\
    All sources & 281,620 & 0.054 & 0.053 & 0.055 \\
    Selected sources & 281,620 & 0.045 & 0.044 & 0.046 \\
    Temporal, all & 281,620 & 0.030 & 0.029 & 0.031 \\
    Temporal, minute precision & 237,465 & 0.033 & 0.031 & 0.034 \\
    Source $\times$ day & 281,620 & 0.206 & 0.205 & 0.207 \\
    Source $\times$ weekday/quarter & 281,620 & 0.418 & 0.417 & 0.419 \\
    Source-matched daily & 281,620 & 0.144 & 0.143 & 0.145 \\
    \midrule
    \multicolumn{5}{l}{\textit{Equal-Count Native Windows}} \\
    All sources & 281,620 & 0.127 & 0.125 & 0.128 \\
    Selected sources & 281,620 & 0.100 & 0.098 & 0.102 \\
    \bottomrule
  \end{tabular*}
\end{table}

The date-method counts are given in Table~\ref{tab:date-extraction}.
JSON-LD supplied 175,104 timestamps (62.18\%), and HTML metadata
supplied 56,904 (20.21\%). Overall, 237,465 records (84.32\%) had
minute-level precision and 44,155 (15.68\%) had day-only precision.
Every published record had a parseable capture timestamp; its stored
publication time lay inside the target interval and rounded down to its
assigned 15-minute bucket.

Native publication-to-batch lags and reconstructed archive capture lags
are examined in Appendix~\ref{app:processing-delay}.

\begin{table}[H]
  \caption{Date Extraction}
  \label{tab:date-extraction}
  \centering
  \footnotesize
  \setlength{\tabcolsep}{3pt}
  \begin{tabular*}{\columnwidth}{@{\extracolsep{\fill}}lrr@{}}
    \toprule
    Method & Minute & Day only \\
    \midrule
    JSON-LD & 172,942 & 2,162 \\
    HTML metadata & 55,288 & 1,616 \\
    Labeled time element & 4,452 & 501 \\
    First time element & 4,783 & 177 \\
    Extractor date fallback & 0 & 39,699 \\
    \bottomrule
  \end{tabular*}
  \ArticleNote{Counts cover all 281,620 published records. The first time element and extractor date fallback are separate heuristic paths. No record used capture time as its publication-date fallback. Precision describes the recorded label, not independently verified accuracy.}
\end{table}

\section{Extended Distribution Comparisons}\label{app:distributions}
This appendix extends the recovery assessment in
Section~\ref{sec:how-much-was-recovered} with detailed comparisons of
source composition and temporal coverage.

Tables~\ref{tab:distribution-comparison}--\ref{tab:frequency-tiers}
report extended distances, the largest source-share errors, and coverage
by native-frequency tier.

Jensen--Shannon divergence (JSD)~\cite{lin1991divergence} is

\begin{equation}
    \begin{aligned}
        \operatorname{JSD}(p, q) &= \tfrac12\sum_i p_i\log_2(p_i/m_i) \\
        &\quad + \tfrac12\sum_i q_i\log_2(q_i/m_i),
    \end{aligned}
\end{equation}
where $p_i$ and $q_i$ are the shares of the reference and reconstructed
distributions in the same category $i$, with $\sum_i p_i=\sum_i q_i=1$,
and $m_i=(p_i+q_i)/2$. Zero-probability terms contribute zero. Base-2
logarithms give values in bits, ranging from zero to one.

Tables~\ref{tab:largest-source-errors} and~\ref{tab:frequency-tiers} use
the full native source frame. In Table~\ref{tab:largest-source-errors},
$\Delta$ is the output share minus the native share, and each source's
TV contribution is $|\Delta|/2$, in percentage points. These
contributions sum to the full-frame TV when taken over all sources. In
Table~\ref{tab:frequency-tiers}, Present is the percentage of sources
with at least one output record. Median attained volume includes
zero-output sources as zeros; Native and Output give each tier's share
of the corresponding record total.

\begin{table}[!t]
  \caption{Distribution Comparisons}
  \label{tab:distribution-comparison}
  \centering
  \footnotesize
  \setlength{\tabcolsep}{3pt}
  \begin{tabular*}{\columnwidth}{@{\extracolsep{\fill}}lrrr@{}}
    \toprule
    Comparison & TV & JSD & Ideal TV \\
    \midrule
    \multicolumn{4}{l}{\textit{Source Shares}} \\
    All native sources & 0.382 & 0.196 & 0.054 \\
    Selected sources & 0.358 & 0.175 & 0.045 \\
    \midrule
    \multicolumn{4}{l}{\textit{Temporal Profile}} \\
    All records & 0.244 & 0.091 & 0.030 \\
    Minute precision & 0.171 & 0.035 & 0.033 \\
    \midrule
    \multicolumn{4}{l}{\textit{Sources Over Time}} \\
    Source $\times$ day & 0.503 & 0.326 & 0.206 \\
    Source $\times$ weekday/quarter & 0.861 & 0.806 & 0.418 \\
    Source-matched daily shape & 0.292 & 0.143 & 0.144 \\
    \bottomrule
  \end{tabular*}
  \ArticleNote{Ideal TV is the mean simulated TV (Section~\ref{sec:comparison-metrics}).}
\end{table}

\begin{table}[!t]
  \caption{Largest Source-Share Differences}
  \label{tab:largest-source-errors}
  \centering
  \footnotesize
  \setlength{\tabcolsep}{3pt}
  \begin{tabular*}{\columnwidth}{@{\extracolsep{\fill}}lrrrr@{}}
    \toprule
    Source & \shortstack{Native (\%)} & \shortstack{Output (\%)} & $\Delta$ (pp) & TV (pp) \\
    \midrule
    \texttt{iheart.com} & 6.04 & 8.63 & +2.59 & 1.30 \\
    \texttt{yahoo.com} & 3.28 & 4.03 & +0.75 & 0.38 \\
    \texttt{marketscreener.com} & 0.43 & 0.00 & -0.43 & 0.22 \\
    \texttt{dailymail.co.uk} & 0.39 & 0.00 & -0.39 & 0.20 \\
    \texttt{boredpanda.com} & 0.41 & 0.06 & -0.35 & 0.18 \\
    \texttt{tvguide.co.uk} & 1.56 & 1.22 & -0.34 & 0.17 \\
    \texttt{themarketsdaily.com} & 0.35 & 0.01 & -0.34 & 0.17 \\
    \texttt{dailypolitical.com} & 0.34 & 0.00 & -0.33 & 0.17 \\
    \texttt{openpr.com} & 0.32 & 0.01 & -0.31 & 0.16 \\
    \texttt{tickerreport.com} & 0.31 & 0.00 & -0.31 & 0.16 \\
    \bottomrule
  \end{tabular*}
\end{table}

\begin{table}[!t]
  \caption{Coverage by Source Frequency}
  \label{tab:frequency-tiers}
  \centering
  \footnotesize
  \setlength{\tabcolsep}{3pt}
  \begin{tabular*}{\columnwidth}{@{\extracolsep{\fill}}lrrrrr@{}}
    \toprule
    Rank & Sources & \shortstack{Present\\(\%)} & \shortstack{Native\\(\%)} & \shortstack{Output\\(\%)} & \shortstack{Median $A_s$\\(\%)} \\
    \midrule
    1--10 & 10 & 80.00 & 14.99 & 16.45 & 7.90 \\
    11--100 & 90 & 65.56 & 13.05 & 8.67 & 3.22 \\
    101--1,000 & 900 & 88.00 & 33.46 & 40.04 & 23.65 \\
    1,001+ & 11,459 & 30.04 & 38.49 & 34.84 & 0.00 \\
    \bottomrule
  \end{tabular*}
  \ArticleNote{Attained-volume ratio $A_s$ is defined in (\ref{eq:attained-volume}).}
\end{table}

On the source-by-weekday/quarter grid, TV is 0.861, compared with an
ideal-sampling mean of 0.418. The grid has 672 temporal positions per
source, with only two or three native observations per position.
Moreover, 65.54\% of reconstructed mass falls in cells with zero native
reference counts, compared with 0.51\% on the source-by-day grid. These
support limitations make the finer grid difficult to interpret as a
measure of temporal agreement.

\begin{figure}[!b]
  \centering
  \includegraphics[width=\columnwidth]{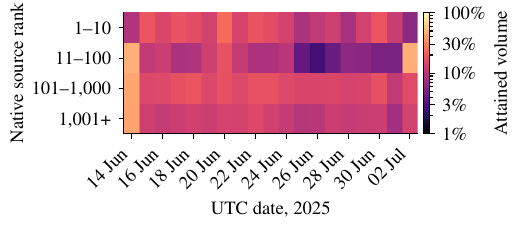}
  \caption{Daily attained volume by native source-frequency tier.
  Each cell divides reconstructed records by the unscaled native
  estimate for the same tier and included target intervals.
  Colors use a logarithmic ratio scale, without normalization to the
  reconstructed total. The first and last dates are partial.}
  \label{fig:source-tier-time}
\end{figure}

\begin{figure}[!t]
\centering
\includegraphics[width=\columnwidth]{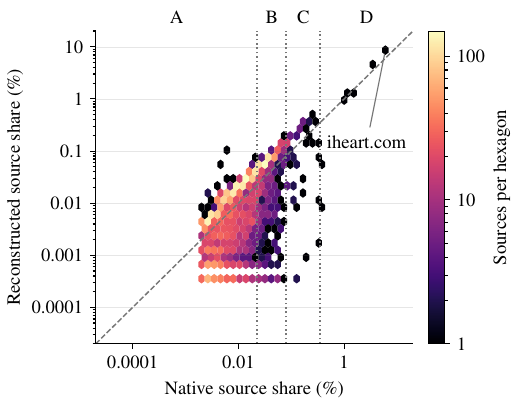}
\caption{Native and reconstructed source shares among represented sources on logarithmic axes. Color gives the number of sources per hexagon on a logarithmic scale. Vertical dotted lines delimit native-frequency ranks: A, 1,001+; B, 101--1,000; C, 11--100; D, 1--10. The dashed line denotes equality. Horizontal shelves arise from integer reconstructed counts.}
\label{fig:source-shares-hexbin}
\end{figure}

\begin{figure}[!t]
\centering
\includegraphics[width=\columnwidth]{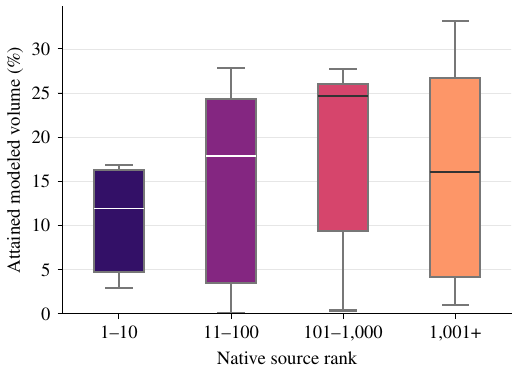}
\caption{Attained modeled volume by native-frequency tier among represented sources. Boxes span the interquartile range, with median lines and whiskers at the 5th and 95th percentiles. Table~\ref{tab:frequency-tiers} includes zero-output sources in its tier medians.}
\label{fig:source-attainment-tiers}
\end{figure}

Fig.~\ref{fig:source-tier-time} shows daily attained volume by frequency
tier. Lower attainment for ranks 11--100 than 101--1,000 on most dates
indicates a persistent tier imbalance.
Figs.~\ref{fig:source-shares-hexbin}
and~\ref{fig:source-attainment-tiers} compare source shares and
tier-attainment distributions, respectively.

\section{Intact-Week Recovery Details}\label{app:holdout}
The 5,000-article source-balanced subset uses a capacity-limited integer
allocation against the outside-week source shares, followed by seed-42
URL-hash selection within sources. Native overlap and field scores are
not selection criteria.

The primary document comparison is exact source--URL intersection. A
sensitivity analysis normalizes host case, default ports, and fragments
while preserving scheme, path, and query. The native batch frame and the
tagged publication-time frame have separate denominators. The latter
uses valid publication tags no later than batch time in observed files
from 1--14 June. Missing tags and boundary crossings are reported
separately.

Paired field scores use one native and one reconstructed observation per
source--URL pair. We score the records with the \software\ 2.2.0
benchmark API, using its strict parsers and within-document set matching
for all observed themes, normalized person and organization identities,
and location identities
$(\mathrm{type},\mathrm{country},\mathrm{ADM1},\mathrm{feature\ ID})$.
Numeric tone agreement includes mean absolute error (MAE), bias, and
concordance correlation coefficient (CCC). Malformed native fields are
excluded from their field and audited. The English-source subset permits
comparison with the processor study~\cite{klimin2026agreement}, which
evaluated a different article population. A title/length diagnostic
requires equality after HTML unescaping, case folding and whitespace
normalization, and a reconstructed/native word-count ratio within
$[0.90,1.10]$ with a positive native denominator. The processor study
used Unicode/token normalization and title-eligibility filters, so these
diagnostic subsets are not identically defined.

The unprocessed acquisition pool contains 5,919 native batch-frame
matches among 15,329 source--URL observations (38.61\%). After text and
language preparation, the full eligible pool contains 5,918 exact
matches; conservative URL normalization gives 6,058. URLs outside the
observed native set are not classified as acquisition errors: native GKG
is an observed reference, not a census of all publications.

Table~\ref{tab:holdout-clocks} compares matching under batch and
publication-time frames using their respective denominators, while
Table~\ref{tab:holdout-tiers} reports URL coverage by reference source
rank.

\begin{table}[H]
\caption{Sensitivity to URL and Time Frames}\label{tab:holdout-clocks}
\centering\footnotesize\setlength{\tabcolsep}{3pt}
\begin{tabular*}{\columnwidth}{@{\extracolsep{\fill}}lrrr@{}}
\toprule Population & \shortstack{Publication\\matches} & \shortstack{Coverage\\(\%)} & \shortstack{Normalized\\matches} \\ \midrule
Full pool & 3,432 & 0.731 & 6,058 \\
Balanced subset & 1,314 & 0.280 & 2,101 \\
\bottomrule\end{tabular*}
\ArticleNote{The publication frame contains 469,740 unique source--URL pairs with parsed nonnegative-lag publication tags in observed 1--14 June files. Normalized matches use the native batch frame and conservative URL normalization.}\end{table}

\begin{table*}[!t]
\caption{Full-Pool URL Coverage by Reference Source Rank}\label{tab:holdout-tiers}
\centering\footnotesize\setlength{\tabcolsep}{3pt}
\begin{tabular*}{\textwidth}{@{\extracolsep{\fill}}lrrrr@{}}
\toprule Reference rank & Sources & Native URLs & Matches & Coverage (\%) \\ \midrule
1--10 & 10 & 127,612 & 369 & 0.289 \\
11--100 & 90 & 110,403 & 246 & 0.223 \\
101--1000 & 900 & 279,031 & 1,610 & 0.577 \\
1001+ & 4,514 & 297,765 & 3,693 & 1.240 \\
Outside selection & 4,781 & 31,137 & 0 & 0.000 \\
\bottomrule\end{tabular*}
\ArticleNote{Ranks use the outside-week source reference. Sources are those observed in the native/reconstructed union; selected sources absent from both frames do not appear. Coverage divides exact matches by native unique source--URL pairs.}\end{table*}

\begin{table*}[!t]
\caption{Intact-Week Content Measures}\label{tab:holdout-content}
\centering\footnotesize\setlength{\tabcolsep}{3pt}
\begin{tabular*}{\textwidth}{@{\extracolsep{\fill}}lrrr@{}}
\toprule Measure & Reference model & Native week & Recovered \\ \midrule
Mean tone & -0.535 & -0.629 & 0.140 \\
ARMEDCONFLICT (\%) & 11.50 & 11.22 & 7.99 \\
MILITARY (\%) & 4.99 & 4.29 & 7.52 \\
Iran country mention (\%) & 1.69 & 1.26 & 1.17 \\
\bottomrule\end{tabular*}
\ArticleNote{Theme and country shares use articles as denominators and can overlap. Reference records are weighted by reciprocal weekly-position exposure; native and recovered columns use their observed articles. These are pooled content measures, without source standardization.}\end{table*}

The paired full pool spans 1,721 sources. Its 27 non-English-source
pairs are too few for a language-specific comparison. English-only
scores in Table~\ref{tab:holdout-agreement} use 5,891 pairs from 1,710
sources. Micro F1 pools within-document true positives, false positives,
and false negatives; both-empty fields contribute no items. The strict
package location identity includes administrative level 1 as well as
type, country and feature ID. The location parser excludes one malformed
native field in the full and English-only populations, leaving 5,917 and
5,890 usable location pairs, respectively.

Table~\ref{tab:holdout-content} compares article-presence shares and
tone across the exposure-weighted reference, observed native week, and
full recovered pool. The reference uses 2,403,281 observed records
across the 21 reference calendar dates, with absent batches excluded.
Inverse weekly-position exposure gives a modeled one-week record mass of
835,279.8 over all reference sources. The selected-source volume model
used for acquisition is 800,914 records; these are different source
frames. Content shares use article denominators and are not normalized
to sum to one across labels.

\section{Publication Time and Processing Delay}\label{app:processing-delay}
Reconstructed buckets use extracted publication time, while native GKG
batch labels describe a later processing stage. Comparisons across the
outage therefore involve different clocks. Native publication-to-batch
lags and reconstructed archive capture lags describe these stages but do
not establish historical first availability.

All 2,490,627 valid native \texttt{DATE} values matched their record-ID
batch prefix. Valid precise-publication tags were present in 1,460,974
records (58.66\%). Within that subset, 221,871 records (15.19\%) fell on
another UTC day and 215,727 (14.77\%) on another weekday. The median
batch-to-publication lag was 129 minutes, with an interquartile range of
57--240 minutes. A further 56,144 publication tags (3.84\%) followed
their batch time. The two native weekday-share distributions had TV of
0.018. The native publication field had no per-record precision label,
and 23,284 tags were at 00:00 UTC; precision for these midnight values
was unspecified.

The median lag was 128 minutes before the gap and 130 minutes after it;
the respective upper quartiles were 239 and 242 minutes. Among 1,404,830
nonnegative lags, 75.82\% exceeded one hour, 54.00\% exceeded two hours,
and 25.57\% exceeded four hours. The pooled 95th percentile over all
tagged records was 1,207 minutes.

\begin{figure}[!t]
\centering
\includegraphics[width=0.86\columnwidth]{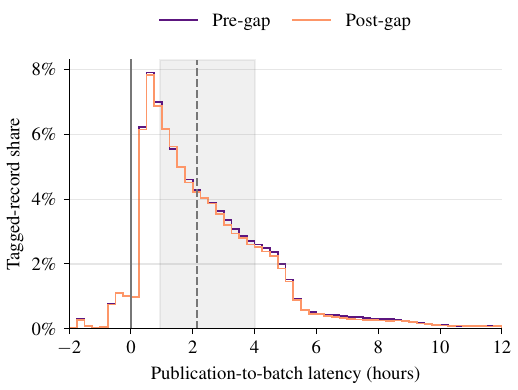}
\caption{Native publication-to-batch lag before and after the outage,
in 15-minute bins. Shading marks the pooled IQR; the dashed line marks
the pooled median.}
\label{fig:native-latency-distribution}\end{figure}

\begin{figure}[!t]
\centering
\includegraphics[width=0.86\columnwidth]{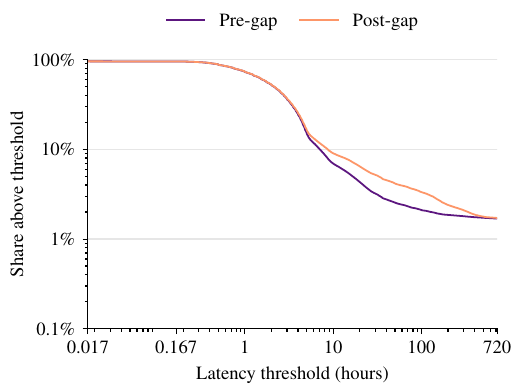}
\caption{Fraction of native publication tags exceeding each lag
threshold before and after the outage, on logarithmic axes.}
\label{fig:native-latency-tail}\end{figure}

Fig.~\ref{fig:native-latency-distribution} shows the native
publication-to-batch lag distribution for 1--14 June and 2--8 July 2025.
Both native-lag figures use all valid publication tags in each period as
their denominators, including negative lags.
Fig.~\ref{fig:native-latency-tail} shows the fraction of native
publication tags exceeding each lag threshold, from one minute to 30
days; 24,848 pooled lags exceed 30 days. The long right tail means that
a fixed shift near the median would still leave the most delayed tagged
records far from their native batch times.

For reconstructed articles, archive capture time supplies a separate lag
measure. Among minute-precision records, median capture lag was 5.71
hours for Wayback and 1.15 hours for Common Crawl.
Table~\ref{tab:capture-lag} reports capture-minus-publication lag for
both precision groups. Minute-precision lags use the stored timestamp;
day-only dates define a 24-hour publication interval and hence bounds on
each lag summary. A negative lower bound alone is compatible with
same-day capture, whereas negative point lags or upper bounds place
capture before the recorded publication interval. In 8,062 records,
capture preceded the stored publication interval. These conflicts can
arise from date extraction or page changes and do not establish that an
article was available before publication.

A hypothetical forward offset changes the temporal comparison and can
move records across the outage boundaries. A fixed offset cannot
reproduce heterogeneous native publication-to-batch lags, and day-only
dates do not supply an intraday placement.
Table~\ref{tab:latency-scenarios} reports sensitivity scenarios for
minute-precision records. Later is the percentage of the 1,404,830
nonnegative native lags exceeding the offset; Outside counts shifted
records beyond the target interval. TV compares the shifted profile
within that interval with the native weekly reference, normalizing each
profile separately. The released dataset preserves extracted publication
timestamps without any delay adjustment; offsets are used only in these
sensitivity calculations.

A two-hour offset gives temporal TV of 0.161, compared with 0.171
without an offset, but moves 1,787 records beyond the target endpoint
and leaves 1,657 occupied target intervals. Four- and six-hour offsets
give TV of 0.180 and 0.211. Shifting the publication-selected collection
also cannot supply articles published before the outage that native
processing might have assigned to its first batches. The scenarios
illustrate clock and boundary sensitivity; they do not reconstruct
historical availability or define a preferred timestamp correction.

\begingroup
\setlength{\intextsep}{6pt}
\begin{table}[H]
  \caption{Archive Capture Lag}
  \label{tab:capture-lag}
  \centering
  \footnotesize
  \setlength{\tabcolsep}{3pt}
  \begin{tabular*}{\columnwidth}{@{\extracolsep{\fill}}lrrr@{}}
    \toprule
    Precision & Records & \shortstack{Median (days)} & \shortstack{95th (days)} \\
    \midrule
    \multicolumn{4}{l}{\textit{Common Crawl}} \\
    Day only & 16,174 & $[-0.39,0.61]$ & $[2.39,3.39]$ \\
    Minute & 8,829 & 0.048 & 1.124 \\
    \midrule
    \multicolumn{4}{l}{\textit{Wayback}} \\
    Day only & 27,981 & $[-0.11,0.89]$ & $[4.54,5.54]$ \\
    Minute & 228,636 & 0.238 & 2.674 \\
    \bottomrule
  \end{tabular*}
  \ArticleNote{Brackets give bounds from day-only precision, not confidence intervals.}
\end{table}

\begin{table}[H]
  \caption{Processing-Delay Scenarios}
  \label{tab:latency-scenarios}
  \centering
  \footnotesize
  \setlength{\tabcolsep}{3pt}
  \begin{tabular*}{\columnwidth}{@{\extracolsep{\fill}}lrrr@{}}
    \toprule
    Offset (h) & Later (\%) & Outside & TV \\
    \midrule
    0 & 99.8 & 0 & 0.171 \\
    1 & 75.8 & 875 & 0.164 \\
    2 & 54.0 & 1,787 & 0.161 \\
    4 & 25.6 & 2,375 & 0.180 \\
    6 & 12.7 & 3,165 & 0.211 \\
    \bottomrule
  \end{tabular*}
  \ArticleNote{Later: native nonnegative lags exceeding the offset. Outside: reconstructed records shifted beyond the target interval.}
\end{table}

\newpage
\endgroup

\ArticleReferences

@inproceedings{barbaresi2021trafilatura,
  author    = {Barbaresi, Adrien},
  title     = {{Trafilatura}: A web scraping library and command-line tool for text discovery and extraction},
  booktitle = {Proceedings of the 59th Annual Meeting of the Association for Computational Linguistics and the 11th International Joint Conference on Natural Language Processing: System Demonstrations},
  publisher = {Association for Computational Linguistics},
  pages     = {122--131},
  month     = aug,
  year      = {2021},
  doi       = {10.18653/v1/2021.acl-demo.15}
}

@misc{klimin2026gdeltregkg,
  author       = {Klimin, Nikita A.},
  title        = {{gdelt-regkg}},
  day          = {29},
  month        = oct,
  year         = {2026},
  howpublished = {Computer software},
  publisher    = {Python Package Index},
  version      = {2.2.0},
  url          = {https://pypi.org/project/gdelt-regkg/2.2.0/}
}

@techreport{leetaru2015gkgcodebook,
  author      = {Leetaru, Kalev},
  title       = {The {GDELT} Global Knowledge Graph ({GKG}) data format codebook {V2.1}},
  year        = {2015},
  month       = feb,
  day         = {19},
  version     = {2.1},
  institution = {{The GDELT Project}},
  type        = {Data format codebook},
  url         = {https://data.gdeltproject.org/documentation/GDELT-Global_Knowledge_Graph_Codebook-V2.1.pdf}
}

@misc{stahl2026lingua,
  author       = {Stahl, Peter M.},
  title        = {Lingua},
  day          = {9},
  month        = mar,
  year         = {2026},
  howpublished = {Computer software},
  publisher    = {PyPI},
  version      = {2.2.0},
  url          = {https://pypi.org/project/lingua-language-detector/2.2.0/}
}

@article{fan2021m2m100,
  author  = {Fan, Angela and Bhosale, Shruti and Schwenk, Holger and Ma, Zhiyi and El-Kishky, Ahmed and Goyal, Siddharth and Baines, Mandeep and Celebi, Onur and Wenzek, Guillaume and Chaudhary, Vishrav and Goyal, Naman and Birch, Tom and Liptchinsky, Vitaliy and Edunov, Sergey and Grave, Edouard and Auli, Michael and Joulin, Armand},
  title   = {Beyond {English}-Centric Multilingual Machine Translation},
  journal = {Journal of Machine Learning Research},
  volume  = {22},
  number  = {107},
  pages   = {1--48},
  day     = {21},
  month   = mar,
  year    = {2021},
  url     = {https://jmlr.org/papers/v22/20-1307.html}
}

@techreport{chapekis2024disappears,
  author      = {Chapekis, Athena and Bestvater, Samuel and Remy, Emma and Rivero, Gonzalo},
  title       = {When Online Content Disappears},
  institution = {Pew Research Center},
  type        = {Research Report},
  year        = {2024},
  month       = may,
  day         = {17},
  url         = {https://www.pewresearch.org/data-labs/2024/05/17/when-online-content-disappears/}
}

@misc{klimin2026agreement,
  author        = {Klimin, Nikita A.},
  title         = {Agreement Between Reconstructed and Native {GDELT} {GKG} Records: A Field-Level Baseline},
  month         = oct,
  year          = {2026},
  eprint        = {2610.35358},
  archiveprefix = {arXiv},
  primaryclass  = {cs.CL},
  doi           = {10.48550/arXiv.2610.35358},
  url           = {https://arxiv.org/abs/2610.35358}
}

@online{eventregistryplans,
  author  = {{Event Registry}},
  title   = {{News API} Pricing},
  url     = {https://newsapi.ai/plans},
  urldate = {2026-10-28}
}

@online{newsdataarchive,
  author  = {{NewsData.io}},
  title   = {News Archive},
  url     = {https://newsdata.io/documentation/news-archive-api},
  urldate = {2026-10-28}
}

@online{newsdatapricing,
  author  = {{NewsData.io}},
  title   = {Pricing Plan by {NewsData.io}},
  url     = {https://newsdata.io/pricing},
  urldate = {2026-10-28}
}

@online{newsdataacademic,
  author  = {{NewsData.io}},
  title   = {Student and University Discount},
  url     = {https://newsdata.io/get-discount},
  urldate = {2026-10-29}
}

@online{proquestfactivacontent,
  author  = {{ProQuest}},
  title   = {{Dow Jones Factiva}: Content},
  url     = {https://proquest.libguides.com/factiva/content},
  urldate = {2026-10-28}
}

@online{lexisnexisnewsdata,
  author  = {{LexisNexis}},
  title   = {News \& Media Data},
  url     = {https://www.lexisnexis.com/en-int/products/nexis-data/data-sets/news-media},
  urldate = {2026-10-28}
}

@online{mediacloudfaq,
  author  = {{Media Cloud}},
  title   = {{Media Cloud} {FAQs}},
  url     = {https://www.mediacloud.org/documentation/faqs},
  urldate = {2026-10-28}
}

@online{nowcorpusdata,
  author  = {{English-Corpora.org}},
  title   = {Full-Text Corpus Data: {NOW} Corpus},
  url     = {https://www.corpusdata.org/now_corpus.asp},
  urldate = {2026-10-28}
}

@online{guardianaccess,
  author  = {{The Guardian}},
  title   = {Open Platform: Get Started},
  url     = {https://open-platform.theguardian.com/access/},
  urldate = {2026-10-28}
}

@online{nytimesarchiveapi,
  author  = {{The New York Times}},
  title   = {The Archive {API}},
  url     = {https://github.com/nytimes/public_api_specs/blob/master/archive_api/archive_api.md},
  urldate = {2026-10-28}
}

@online{commoncrawlfaq,
  author  = {{Common Crawl}},
  title   = {Frequently Asked Questions},
  url     = {https://commoncrawl.org/faq},
  urldate = {2026-10-27}
}

@online{internetarchivecdxapi,
  author  = {{Internet Archive}},
  title   = {Wayback {CDX} Server {API}},
  url     = {https://github.com/internetarchive/wayback/tree/master/wayback-cdx-server},
  urldate = {2026-10-27}
}

@online{commoncrawl2024terms,
  author  = {{Common Crawl}},
  title   = {Terms of Use},
  day     = {7},
  month   = mar,
  year    = {2024},
  url     = {https://commoncrawl.org/terms-of-use},
  urldate = {2026-10-27}
}

@inproceedings{braun2026validating,
  author    = {Braun, Laura and Oswald, Christian},
  title     = {Validating a Pipeline to Create a Comparable Corpus of Government-Issued Travel Advisories from the {Internet Archives}},
  booktitle = {Proceedings of the 19th Workshop on Building and Using Comparable Corpora (BUCC)},
  publisher = {ELRA Language Resources Association},
  address   = {Palma de Mallorca, Spain},
  pages     = {96--107},
  month     = may,
  year      = {2026},
  doi       = {10.63317/5c34j7cbnshd},
  url       = {https://aclanthology.org/2026.bucc-1.11/}
}

@online{nagel2016ccnews,
  author       = {Nagel, Sebastian},
  title        = {News Dataset Available},
  organization = {Common Crawl},
  year         = {2016},
  month        = oct,
  day          = {4},
  url          = {https://commoncrawl.org/blog/news-dataset-available},
  urldate      = {2026-10-28}
}

@online{internetarchive2001wayback,
  author       = {{Internet Archive}},
  title        = {{Internet Archive} Launches {Wayback Machine}: Free Service Enables Users to Access Archived Versions of Web Sites Dating from 1996},
  howpublished = {Press release},
  year         = {2001},
  month        = oct,
  day          = {24},
  url          = {https://librarytechnology.org/pr/9384/internet-archive-launches-wayback-machine-free-service-enables-users-to-access-archived-versions-of-web-sites-dating-from-1996},
  urldate      = {2026-10-28}
}

@inproceedings{bermejo2026mediacloud,
  author    = {Bermejo, Fernando and Bhargava, Rahul and Budne, Phil and Gulley, Paige and Leon, Evan and McGrady, Ryan and Ndulue, Emily Boardman and Zuckerman, Ethan},
  title     = {{Media Cloud} 2.0: An Updated Open Web News Archive},
  booktitle = {Proceedings of the International AAAI Conference on Web and Social Media},
  volume    = {20},
  number    = {1},
  pages     = {2735--2746},
  year      = {2026},
  month     = may,
  day       = {25},
  publisher = {AAAI Press},
  doi       = {10.1609/icwsm.v20i1.42778},
  url       = {https://ojs.aaai.org/index.php/ICWSM/article/view/42778}
}

@article{ulloa2025timedelays,
  author  = {Ulloa, Roberto and Mangold, Frank and Schmidt, Felix and Gilsbach, Judith and Stier, Sebastian},
  title   = {Beyond Time Delays: How Web Scraping Distorts Measures of Online News Consumption},
  journal = {Communication Methods and Measures},
  volume  = {19},
  number  = {3},
  pages   = {179--200},
  year    = {2025},
  month   = mar,
  day     = {26},
  doi     = {10.1080/19312458.2025.2482538}
}

@article{sen2021totalerror,
  author  = {Sen, Indira and Fl{\"o}ck, Fabian and Weller, Katrin and Wei{\ss}, Bernd and Wagner, Claudia},
  title   = {A Total Error Framework for Digital Traces of Human Behavior on Online Platforms},
  journal = {Public Opinion Quarterly},
  volume  = {85},
  number  = {S1},
  pages   = {399--422},
  year    = {2021},
  month   = aug,
  day     = {30},
  doi     = {10.1093/poq/nfab018}
}

@article{grimmer2013text,
  author  = {Grimmer, Justin and Stewart, Brandon M.},
  title   = {Text as Data: The Promise and Pitfalls of Automatic Content Analysis Methods for Political Texts},
  journal = {Political Analysis},
  volume  = {21},
  number  = {3},
  pages   = {267--297},
  year    = {2013},
  month   = jul,
  doi     = {10.1093/pan/mps028}
}

@inproceedings{mackenzie2020ccnews,
  author    = {Mackenzie, Joel and Benham, Rodger and Petri, Matthias and Trippas, Johanne R. and Culpepper, J. Shane and Moffat, Alistair},
  title     = {{CC-News-En}: A Large {English} News Corpus},
  booktitle = {Proceedings of the 29th ACM International Conference on Information \& Knowledge Management},
  publisher = {ACM},
  pages     = {3077--3084},
  year      = {2020},
  month     = oct,
  doi       = {10.1145/3340531.3412762}
}

@inproceedings{bevendorff2023extraction,
  author    = {Bevendorff, Janek and Gupta, Sanket and Kiesel, Johannes and Stein, Benno},
  title     = {An Empirical Comparison of Web Content Extraction Algorithms},
  booktitle = {Proceedings of the 46th International ACM SIGIR Conference on Research and Development in Information Retrieval},
  publisher = {ACM},
  address   = {New York, NY, USA},
  pages     = {2594--2603},
  year      = {2023},
  month     = jul,
  day       = {18},
  doi       = {10.1145/3539618.3591920}
}

@inproceedings{weigle2023replay,
  author    = {Weigle, Michele C. and Nelson, Michael L. and Alam, Sawood and Graham, Mark},
  title     = {Right {HTML}, Wrong {JSON}: Challenges in Replaying Archived Webpages Built with Client-Side Rendering},
  booktitle = {2023 ACM/IEEE Joint Conference on Digital Libraries (JCDL)},
  publisher = {IEEE},
  pages     = {82--92},
  year      = {2023},
  month     = jun,
  doi       = {10.1109/JCDL57899.2023.00022}
}

@online{umdcyberevents,
  author  = {{University of Maryland School of Public Policy}},
  title   = {Cyber Events Database Home},
  url     = {https://gotech.spp.umd.edu/research-impact/publications/cyber-events-database-home},
  urldate = {2026-10-30}
}

@online{google2025servicecontrol,
  author  = {{Google Cloud}},
  title   = {Multiple {GCP} Products Are Experiencing Service Issues},
  year    = {2025},
  month   = jun,
  day     = {13},
  url     = {https://status.cloud.google.com/incidents/ow5i3PPK96RduMcb1SsW},
  urldate = {2026-10-30}
}

@article{weninger2015extraction,
  author  = {Weninger, Tim and Palacios, Rodrigo and Crescenzi, Valter and Gottron, Thomas and Merialdo, Paolo},
  title   = {Web Content Extraction: A Meta-Analysis of Its Past and Thoughts on Its Future},
  journal = {ACM SIGKDD Explorations Newsletter},
  volume  = {17},
  number  = {2},
  pages   = {17--23},
  year    = {2015},
  url     = {https://arxiv.org/abs/1508.04066}
}

@inproceedings{ainsworth2011archived,
  author    = {Ainsworth, Scott G. and AlSum, Ahmed and SalahEldeen, Hany and Weigle, Michele C. and Nelson, Michael L.},
  title     = {How Much of the Web Is Archived?},
  booktitle = {Proceedings of the 11th Annual International ACM/IEEE Joint Conference on Digital Libraries},
  publisher = {ACM},
  pages     = {133--136},
  year      = {2011},
  month     = jun,
  doi       = {10.1145/1998076.1998100}
}

@inproceedings{baack2024commoncrawl,
  author    = {Baack, Stefan},
  title     = {A Critical Analysis of the Largest Source for Generative {AI} Training Data: {Common Crawl}},
  booktitle = {Proceedings of the 2024 ACM Conference on Fairness, Accountability, and Transparency},
  publisher = {ACM},
  pages     = {2199--2208},
  year      = {2024},
  month     = jun,
  day       = {5},
  doi       = {10.1145/3630106.3659033}
}

@misc{kwak2014newsgeography,
  author        = {Kwak, Haewoon and An, Jisun},
  title         = {Understanding News Geography and Major Determinants of Global News Coverage of Disasters},
  year          = {2014},
  month         = oct,
  day           = {14},
  eprint        = {1410.3710},
  archiveprefix = {arXiv},
  primaryclass  = {cs.CY},
  doi           = {10.48550/arXiv.1410.3710}
}

@inproceedings{norregaard2019nelagt,
  author    = {N{\o}rregaard, Jeppe and Horne, Benjamin D. and Adal{\i}, Sibel},
  title     = {{NELA-GT-2018}: A Large Multi-Labelled News Dataset for the Study of Misinformation in News Articles},
  booktitle = {Proceedings of the International AAAI Conference on Web and Social Media},
  publisher = {AAAI Press},
  volume    = {13},
  number    = {1},
  pages     = {630--638},
  year      = {2019},
  month     = jun,
  doi       = {10.1609/icwsm.v13i01.3261}
}

@misc{lazzaroni2026infininewsefficientlyqueryableaccess,
  author        = {Lazzaroni, Ruggero Marino and Lasser, Jana and Solovev, Kirill},
  title         = {{Infini-News}: Efficiently Queryable Access to 1.3 Billion Processed Common Crawl News Articles},
  year          = {2026},
  month         = may,
  day           = {18},
  eprint        = {2605.18337},
  archiveprefix = {arXiv},
  primaryclass  = {cs.CL},
  doi           = {10.48550/arXiv.2605.18337}
}

@misc{solovev2026finenews,
  author    = {Solovev, Kirill},
  title     = {{Fine-News}},
  year      = {2026},
  version   = {2},
  publisher = {Hugging Face},
  url       = {https://huggingface.co/datasets/ksolovev/fine-news/tree/v2}
}

@online{gdelt2016articlemetadata,
  author  = {{The GDELT Project}},
  title   = {New {GKG} 2.0 Article Metadata Fields},
  year    = {2016},
  month   = apr,
  day     = {22},
  url     = {https://blog.gdeltproject.org/new-gkg-2-0-article-metadata-fields/},
  urldate = {2026-10-30}
}

@misc{klimin2026regkgdataset,
  author    = {Klimin, Nikita A.},
  title     = {{ReGKG-2025}},
  year      = {2026},
  publisher = {Zenodo},
  version   = {1.0.0},
  doi       = {10.5281/zenodo.23964265},
  url       = {https://doi.org/10.5281/zenodo.23964265}
}

@techreport{unsc2025middleeast9936,
  author      = {{United Nations Security Council}},
  title       = {{Security Council}, 80th Year: 9936th Meeting},
  institution = {United Nations},
  type        = {Verbatim Meeting Record},
  number      = {S/PV.9936},
  year        = {2025},
  month       = jun,
  day         = {13},
  url         = {https://docs.un.org/en/S/PV.9936},
  urldate     = {2026-11-08}
}

@techreport{unsc2025middleeast9941,
  author      = {{United Nations Security Council}},
  title       = {{Security Council}, 80th Year: 9941st Meeting},
  institution = {United Nations},
  type        = {Verbatim Meeting Record},
  number      = {S/PV.9941},
  year        = {2025},
  month       = jun,
  day         = {22},
  url         = {https://docs.un.org/en/S/PV.9941},
  urldate     = {2026-11-08}
}

@techreport{unsc2025europe9965,
  author      = {{United Nations Security Council}},
  title       = {{Security Council}, 80th Year: 9965th Meeting},
  institution = {United Nations},
  type        = {Verbatim Meeting Record},
  number      = {S/PV.9965},
  year        = {2025},
  month       = jul,
  day         = {25},
  url         = {https://docs.un.org/en/S/PV.9965},
  urldate     = {2026-11-08}
}

@techreport{unsc2025middleeast9944,
  author      = {{United Nations Security Council}},
  title       = {{Security Council}, 80th Year: 9944th Meeting},
  institution = {United Nations},
  type        = {Verbatim Meeting Record},
  number      = {S/PV.9944},
  year        = {2025},
  month       = jun,
  day         = {24},
  url         = {https://docs.un.org/en/S/PV.9944},
  urldate     = {2026-11-07}
}

@online{gdelt2026masterfilelist,
  author  = {{The GDELT Project}},
  title   = {{GDELT} 2.0 Master File List},
  url     = {https://data.gdeltproject.org/gdeltv2/masterfilelist.txt},
  urldate = {2026-10-29}
}

@misc{honnibal2020spacy,
  author       = {Honnibal, Matthew and Montani, Ines and Van Landeghem, Sofie and Boyd, Adriane},
  title        = {{spaCy}: Industrial-strength natural language processing in {Python}},
  year         = {2020},
  howpublished = {Computer software},
  publisher    = {Zenodo},
  version      = {3.8.16},
  doi          = {10.5281/zenodo.1212303},
  url          = {https://doi.org/10.5281/zenodo.1212303}
}

@misc{explosion2024encoreweblg,
  author       = {{Explosion}},
  title        = {{en\_core\_web\_lg}},
  year         = {2024},
  month        = sep,
  day          = {30},
  howpublished = {Language model},
  version      = {3.8.0},
  url          = {https://github.com/explosion/spacy-models/releases/tag/en_core_web_lg-3.8.0}
}

@article{lin1991divergence,
  author  = {Lin, Jianhua},
  title   = {Divergence Measures Based on the {Shannon} Entropy},
  journal = {IEEE Transactions on Information Theory},
  volume  = {37},
  number  = {1},
  pages   = {145--151},
  year    = {1991},
  month   = jan,
  doi     = {10.1109/18.61115}
}

@incollection{hale2017livearchive,
  author    = {Hale, Scott A. and Blank, Grant and Alexander, Victoria D.},
  title     = {Live versus Archive: Comparing a Web Archive to a Population of Web Pages},
  booktitle = {The Web as History: Using Web Archives to Understand the Past and the Present},
  editor    = {Br{\"u}gger, Niels and Schroeder, Ralph},
  publisher = {UCL Press},
  address   = {London},
  pages     = {45--61},
  year      = {2017},
  doi       = {10.2307/j.ctt1mtz55k.8},
  url       = {https://ucldigitalpress.co.uk/Book/Article/45/70/3408/}
}
\end{document}